\documentclass{article}

\usepackage[english]{babel}
\usepackage[T1]{fontenc}
\usepackage[utf8]{inputenc}
\usepackage{graphicx}
\usepackage{secdot}
\sectiondot{section}
\usepackage{wrapfig}
\usepackage{float}
\usepackage[a4paper,bindingoffset=0.2in,left=1in,right=1in,top=1in,bottom=1in,footskip=.25in]{geometry}
\usepackage{amsmath}
\usepackage{amssymb}
\usepackage{extarrows}
\usepackage{arydshln} 
\usepackage{outlines}
\usepackage{todonotes} 
\usepackage{siunitx}
\usepackage{mathtools} 
\usepackage{comment} 
\DeclareSIUnit{\ms}{\meter\per\second}
\DeclareSIUnit{\s}{\second}

\usepackage{mathtools} 

\let\vec\mathbf
\newcommand{\Iavg}{$\overline{I_{th}(w)}$}
\newcommand{\nbr}[1]{\ensuremath{\mathrm{#1}}}
\newcommand{\Nuavg}{$\overline{\nbr{Nu}(w)}$}
\newcommand\avg[1]{\overline{#1}}

\begin{document}
	
\title{Wind Variability and Its Effect on Transmission Line Capacity Estimation}

\author{Nika Mlinarič Hribar\thanks{N. Mlinarič Hribar is with the Jožef Stefan Institute, Ljubljana, Slovenia, and the Jožef Stefan Postgraduate School, Ljubljana, Slovenia (e-mail: nika.mlinaric@ijs.si).} ,%
	\\Matjaž Depolli\thanks{M. Depolli is with the Jožef Stefan Institute, Ljubljana, Slovenia (e-mail: matjaz.depolli@ijs.si).} ,%
	\\Gregor Kosec\thanks{G. Kosec is with the Jožef Stefan Institute, Ljubljana, Slovenia (e-mail: gregor.kosec@ijs.si).
	\\
	\\
	Authors acknowledge the financial support from the Slovenian Research and Innovation Agency (ARIS) research core funding No.\ P2-0095, and the HEDGE-IoT project, which has received funding from the European Union's Horizon Europe research and innovation programme under grant agreement No. 101136216.
	\\
	\\
	This manuscript is a pre-print of a paper submitted to \emph{Energy Conversion and Management}: X and is currently under review.
}}%

\maketitle

\section*{Abstract}

This study investigates the impact of wind velocity averaging on Dynamic Thermal Rating (DTR) calculations. It is based on a high-temporal-resolution (\SI{1}{\second}) wind measurements obtained from a transmission line in Slovenia, Europe. Wind speed and direction variability are analysed, and two averaging methods, namely vector averaging, where velocity is averaged as vector, and hybrid averaging, where speed is averaged as scalar, are employed. DTR calculations are performed on both high-resolution data and averaged data (averaging window \SI{5}{\minute}). It is demonstrated that averaging has a significant effect on both Nusselt number and ampacity, and the effect exhibits a strong angular dependency on the relative angle of the wind to the line. Therefore, two limit cases are studied: in the case of parallel wind, averaged data underestimates the ampacity, and there is a significant amount of cases where the underestimation is larger than 10\%. In the case of perpendicular wind, the two averaging methods affect the results in different ways, but both result in a substantial amount of cases where ampacity is overestimated, potentially leading to unsafe operation. The main takeaway of the study is that averaging wind velocity has a significant impact on DTR results, and special emphasis should be given to the averaging method, as different methods affect the results in different ways.

\section{Introduction}

With the increase in energy demand and clean energy transition, the electricity market is undergoing a significant transformation~\cite{bichler2022electricity}. Transmission system operators (TSOs) are looking at ways to optimise their operation and increase the capacity of their overhead power lines. One of the limiting factors is the temperature of the power line, which should not exceed a critical level to prevent excessive sag and minimise structural degradation~\cite{cimini2013temperature}.
 
The quantity of interest to TSOs is ampacity, i.e. the maximum permissible current at which power lines do not exceed the critical temperature. Historically, ampacity was set to a constant value, often referred to as the \emph{static limit}, determined by a set of unfavourable weather conditions (high ambient temperature and solar radiation paired with low wind)~\cite{a_douglass_review_2019}. While the use of static limit ensures safe operation under most conditions, it also causes the line to be significantly underutilised most of the time, when the weather conditions are favourable. Furthermore, the static conditions have a small chance of being violated~\cite{maksic_cooling_2019}, which introduces risks in the operation.

A possible improvement to a static approach is to monitor the conditions and adjust ampacity dynamically, which is called dynamic thermal rating (DTR)~\cite{a_douglass_review_2019, cigre, ieee}. DTR helps in reducing congestion on power lines and in optimising their utilisation~\cite{irena_dynamic_2020} and is becoming increasingly popular with TSOs around the world. The capacity of the whole line is determined by the line span with the lowest ampacity -- the critical span~\cite{matus2012identification}. The critical span often occurs in conjunction with wind parallel to the line and/or low wind speeds. We argue that in such regimes, existing DTR models may underestimate ampacity, indicating both an opportunity to enhance the accuracy and robustness of these models and a potential to optimise the utilization and efficiency of the power grid.

There are several approaches to DTR~\cite{karimi_2018_dynamic}. This paper will focus on the DTR approach that implements physical models, which calculate the heat balance equation for the line based on the surrounding weather conditions and line current~\cite{cigre, ieee, a_douglass_review_2019}. The models comprise Joule heating as the internal heating mechanism, and solar heating, radiative cooling and convective cooling as the heat exchange mechanisms with the surroundings. There are several studies that propose improvements in these models, from handling the uncertainties~\cite{poli_possible_2019,rashkovska_uncertyinty_2022,chen_secure_2024}, to proposing additional heat terms such as precipitation-driven cooling~\cite{pytlak_modelling_2011,kosec_dynamic_2017}. A key input for these models are the weather conditions, i.e. wind speed and direction, ambient temperature, solar radiation, relative humidity, pressure and rain rate. Notably, wind is a major factor, as it characterises convection, the most important cooling mechanism~\cite{hosek_effect_2011,howington_dynamic_1987}. 

The typical temporal resolution of the input data used in DTR is around \qtyrange{1}{10}{\minute}, and the data is averaged over each sampling interval (we will refer to it as averaging window, or short, window). This resolution is sufficient for most of the parameters, as their rate of change is slow. For wind, however, this time scale is quite long, as it is known that both wind speed and direction can vary significantly over a few minutes. Therefore is reasonable to speak of wind variability within the window. 

Wind speed is typically reported with multiple statistics for each measurement. For instance, maximum wind speed is often associated with wind gusts~\cite{suomi_wind_2018}, a topic that has been relatively well-researched in strong wind regimes, where significant gusts can pose potential hazards~\cite{rafei_analysis_2023}. Another commonly reported statistic is speed deviation, which is related to turbulence intensity~\cite{raichle2009wind}. However, in the context of DTR and critical line spans, the opposite scenario, where wind speeds are low, is of particular interest.

Variability in wind directions is also researched in several fields, however, mostly on long temporal scales or high wind regimes. A study of winds strong enough for energy production from wind turbines~\cite{rott_robust_2018} found that wind direction measurements sampled with \SI{1}{\second} sampling period within \SI{5}{\minute} windows were generally distributed normally, with an average standard deviation of \ang{5.3}. Since wind direction variability depends on the wind speed, these findings are of limited use to our efforts. It is generally accepted that wind variability decreases with an increase in speed~\cite{mahrt_surface_2011,shu_analysis_2024}, or similarly, that strong winds have better-defined direction than low winds~\cite{cigre}. A study of air pollutants spread~\cite{houle_near_2023} found that standard deviation of wind direction within \qtyrange{1}{10}{\minute} windows was up to \ang{15}--\ang{75}, depending on the environment, which is considerable.

This brings us back to averaging. Wind is a vector quantity with speed (amplitude) and direction, which are usually measured separately~\cite{meteo}. It can be averaged as a vector, or each of the parameters can be averaged separately, as scalars. In applications considering strong winds, such as wind measurement in marine buoys~\cite{gilhousen_field_1987,thomas_buoy_2011}, or synoptic scales, such as weather forecasts~\cite{meteo}, both averaging methods usually give similar results. With the lower winds and shorter time scales of interest in this paper, this will not necessarily hold true. In DTR studies known to the authors, no consideration has been given to the wind variability within the windows, and the choice of the averaging method. As the relationship between wind velocity and heat loss due to convective cooling is highly non-linear, we expect the process of averaging to impact the results.

The aim of this study is to analyse whether the temporal resolution of the wind measurements affects the DTR results. We will show that indeed, wind variability has a measurable effect on DTR calculations and that the choice of the averaging method can significantly impact the estimated ampacity. This is especially true in the case where the wind is parallel to the line, where taking into account wind variability increase the ampacity. As this often occurs in the critical span, understanding it and accounting for it in DTR would result in increased capacity of the whole line, which is of practical interest to TSOs.

The rest of the paper is structured as follows: First, we look at the wind measurements for a single location in Slovenia in~\ref{sec:wind-data-n-realistic-conditions} and discuss different averaging methods and the effects on averaging in~\ref{sec:wind-variability-within-the-averaging-window}, where we also introduce a measure for describing the variability in wind speed and direction and analyse the wind variability for the observed location, and how it depends on the average wind speed and direction. We look at how the variability affects the DTR simulation (namely Nusselt number and ampacity) in~\ref{sec:dtr}, and look at its dependence on the wind relative angle in~\ref{sec:angle-dependency}. We present two limit cases with the wind parallel and perpendicular to the conductor in~\ref{sec:limit-cases}, where we demonstrate that averaging has a significant effect on the calculated ampacity. We also include a short analysis of how the window length affects the calculations in~\ref{Appendix}.

\section{Wind data in realistic conditions - acquisition and averaging}
\label{sec:wind-data-n-realistic-conditions}

\begin{figure}[H]
	\begin{center}
		\includegraphics[width=\textwidth]{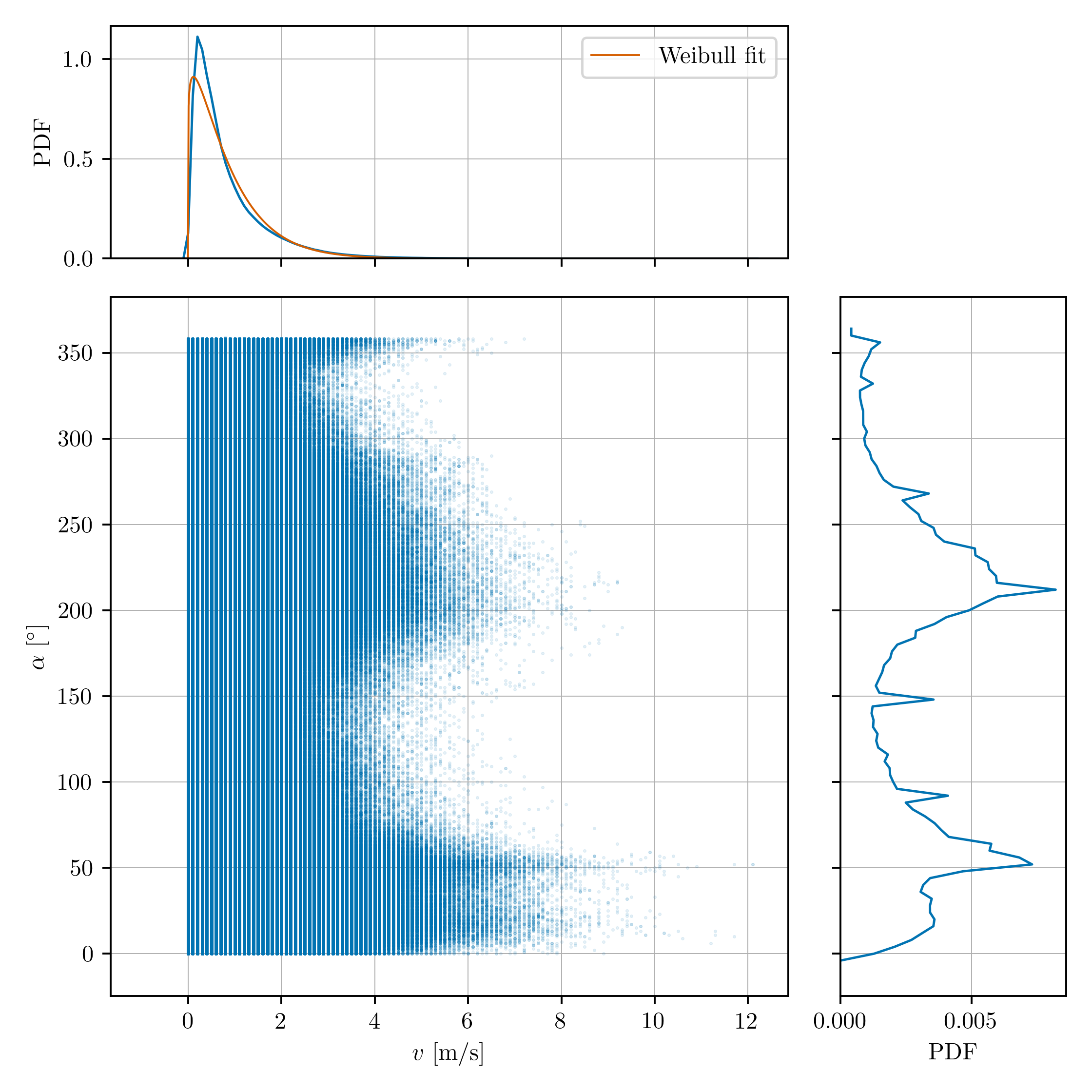}
	\end{center}
	\caption{Scatter plot and PDFs of measured wind data for April, August and September 2024, i.e. observed data. Weibull distribution is fitted on wind speed distribution using maximum likelihood estimation (MLE).}
	\label{fig:wind_vAndAlphaSource}
\end{figure}

For the purposes of this paper, Slovenian TSO ELES provided wind data with a temporal resolution of \SI{1}{\second}, measured with \emph{WXT 536} ultrasonic wind sensor located on \SI{220}{\kilo\volt} Podlog-Obersielach AlFe 240/40 \SI{}{\milli\meter^2} line. The sensor's range for wind speed is \qtyrange[range-units = single]{0}{60}{\ms} with resolution of \SI{0.1}{\ms} and wind angle resolution is \ang{1}. Both quantities are measured with accuracy of \SI{3}{\percent} at \SI{10}{\ms}. The observed data covers two separate periods: from the 1st to the 30th of April and from the 1st of August to the 30th of September 2024. These were the only periods where the high-resolution measurements were available. Figure~\ref{fig:wind_vAndAlphaSource} shows the scatter plot of the measurements together with speed and angle probability density functions (PDFs), where $\vec{v}\left({v,\alpha}\right)$ stands for the measured \SI{1}{\second} data, with $v$ and $\alpha$ denoting the wind speed (magnitude) and angle.

First, we test whether wind speed distribution can be modelled with standard distributions. Weibull distribution is fitted on wind speed distribution, as it is widely used in energy production and transport analysis to model wind speed distributions and long-term wind variability~\cite{hosek_effect_2011, ward_time-averaged_2023, gonzalez-cagigal_influence_2022}. This distribution is particularly useful because it can effectively describe the long-term (daily, monthly, seasonal) variability of wind speed over time.  We observe that the general PDF shape loosely follows the Weibull distribution, however, we reject the Weibull distribution with the Kolmogorov-Smirnov test (p-value $< \mathrm{e}^{-10^4}$). This is expected, as previous studies have found that the Weibull distribution fit is the poorest in cases with a lot of calms and non-circular wind velocity distribution~\cite{tuller-brett_characteristics_1984}, which are both true for the observed data.

\begin{figure}[H]
	\begin{center}
		\includegraphics[width=0.49\textwidth]{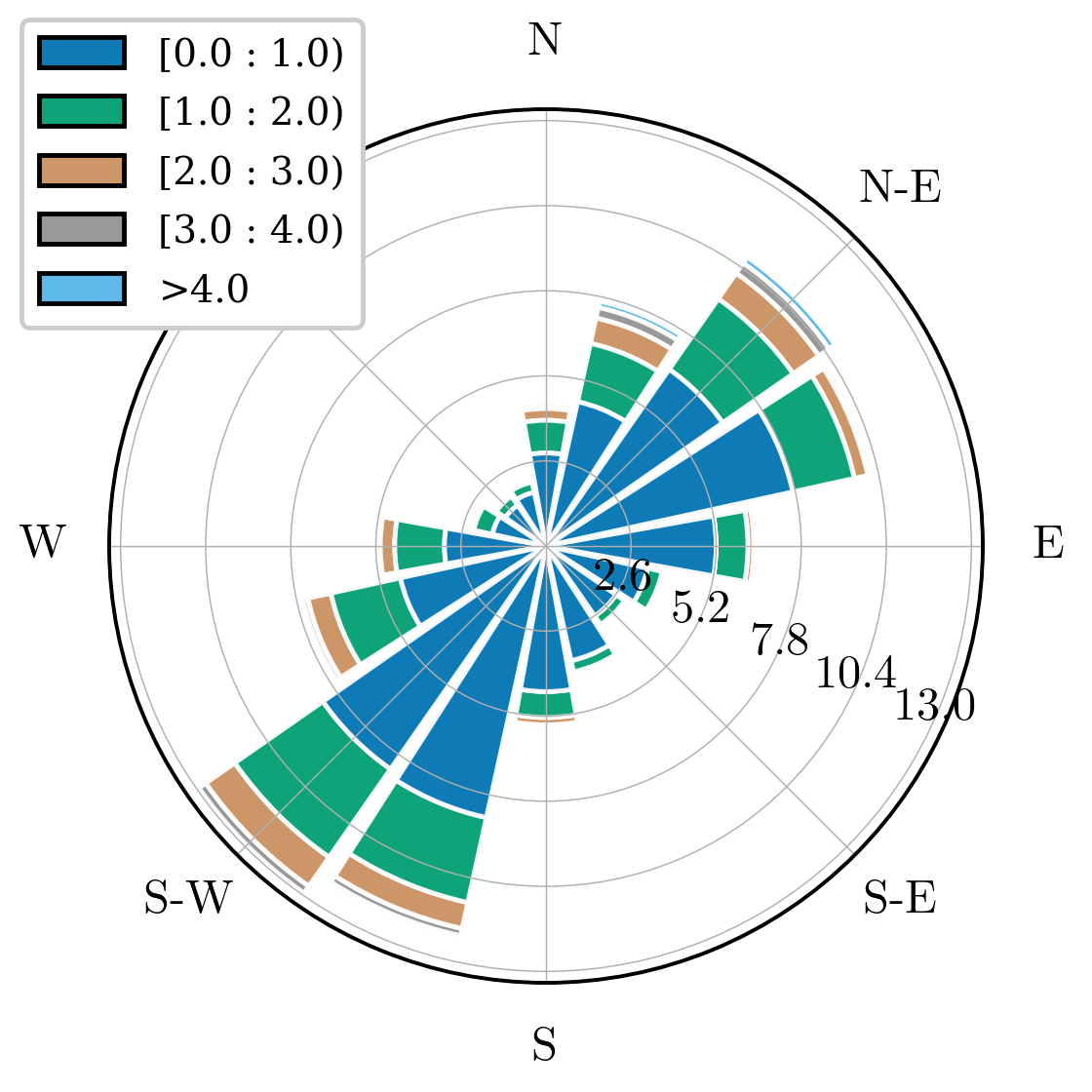}
		\includegraphics[width=0.49\textwidth]{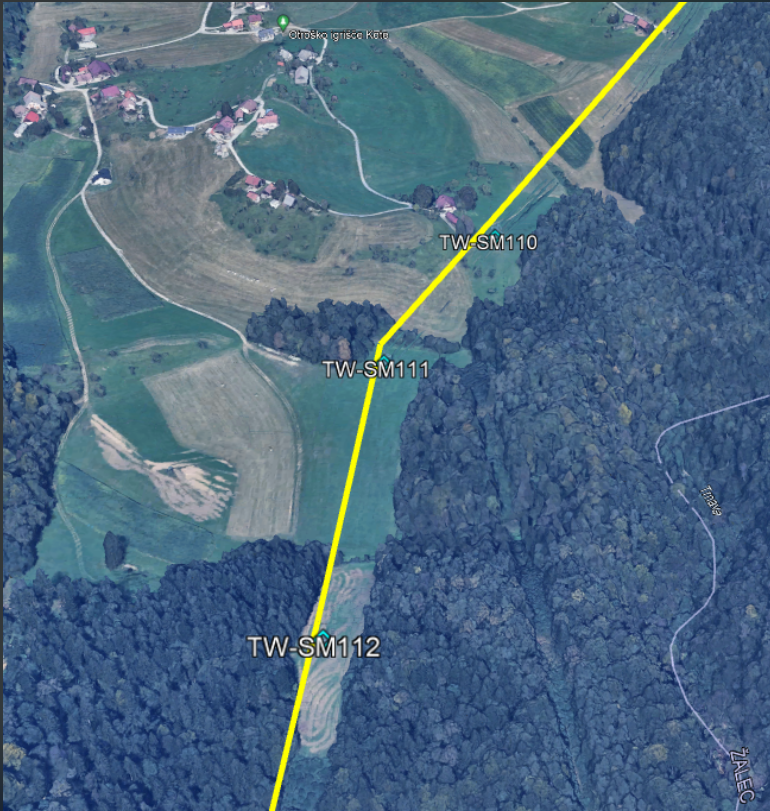}
	\end{center}
	\caption{Wind rose (left) and satellite image of the considered site. The span between pylons SM111 and SM112 will be the subject of DTR computations later in the manuscript.}
	\label{fig:wind_windrose}
\end{figure}

Next, we take a look at the wind direction. There are two predominant wind directions in the wind angle distribution: one at about \ang{55} and the other at about \ang{210}, where most of the higher speed measurements take place. For a better visualisation, refer to the wind rose in Figure~\ref{fig:wind_windrose} (left). A look at the satellite image of the site in Figure~\ref{fig:wind_windrose} (right) reveals that the observed strong directional dependence is likely the consequence of the topographic features.

Using a \SI{1}{\second} time resolution is not practical for most DTR applications. In the case of measurement-based DTR, the primary limitation is often the required data communication bandwidth. On the other hand, the temporal resolution of numerical weather prediction is constrained by both the simulation time step and the underlying physical model. Therefore, DTR systems typically rely on averaged values over a time interval with a predefined length window, which is generally between \SI{5}{\minute} and \SI{10}{\minute}. In~\cite{hosek_effect_2011}, the authors argue that using a window longer than \SI{10}{\minute} can significantly degrade the accuracy of DTR results. Before exploring the effects of averaging, however, it is important to first clarify the concept of averaging itself. Wind, being a vector quantity, is characterised by both magnitude (speed) and direction (angle).

In practice, there are two fundamentally different averaging approaches in use~\cite{meteo,copernicus_how_nodate,gilhousen_field_1987}. The first is vector averaging, where wind vectors are treated as geometric entities, with their resultant representing the average vector. The second is hybrid approach, where each component of the wind velocity is averaged separately. The wind magnitude is averaged as a scalar value, while the average angle is determined using vector averaging, either by taking the wind speed into account, or using a unit vector. For the observed location, the TSO uses vector averaging, however, averaging itself is often not discussed at all when talking about DTR, so we assume TSOs take the default sensor output, and do not concern themselves with the averaging method used by the sensor. In this paper, we will take a look at two approaches: vector averaging, and hybrid averaging where speed is taken into account in direction average.

Mathematically, vector average $\avg{\vec{v}_v}\left( \avg{v_v}, \avg{\alpha_v}\right)$ reads as
\begin{equation}
	\avg{\vec{v}_v} = \frac{1}{n} \sum_{1}^{n}\vec{v}
	\label{eq:def_vec_avg}
\end{equation}
and hybrid average $\avg{\vec{v}_h}\left( \avg{v_h}, \avg{\alpha_h}\right)$ as
\begin{equation}	
	\avg{\vec{v}_h} = \left( \frac{1}{n} \sum_{1}^{n} v, \avg{\alpha_v} \right)
	\label{eq:def_mix_avg}
\end{equation}
with $n$ standing for number of samples in one window. Generally, if the application only depends on wind speed, scalar average is more suitable, and if direction is also important, like in particle transport, vector averaging performs better\cite{copernicus_how_nodate}.

In DTR, both speed and direction are important, and as we will see in Section~\ref{sec:dtr}, where CIGRE mathematical model for convective heat exchange with the surrounding air is discussed, the relationship between wind velocity and heat flux is highly nonlinear. As a result, there is no one obvious averaging method choice, and the averaging methods has a significant impact on DTR. The average value, calculated on high-resolution data and then averaged, differs from the value calculated using averaged inputs.

For instance, consider an extreme case where during half of the window, significant wind blows parallel to the line from one side, and during the other half, it blows from the opposite direction with equal magnitude. The DTR cooling effect would be the same as if the wind had been consistently blowing from one (parallel) direction. However, in such a scenario, vector averaging would yield a net wind of zero (therefore underestimating the cooling), and in hybrid averaging, the wind angle would not be defined. If the wind direction would have a slight deviation from the parallel, however, the average angle would be at \ang{90} relative to the line and we would get a constant perpendicular wind. We will see in section~\ref{sec:dtr} that convective cooling depends on the relative angle of the wind, and that it's most efficient when the wind is perpendicular to the line, which means hybrid averaging would overestimate the cooling in this case.

\begin{figure}[H]
	\begin{center}
		\includegraphics[width=0.49\textwidth]{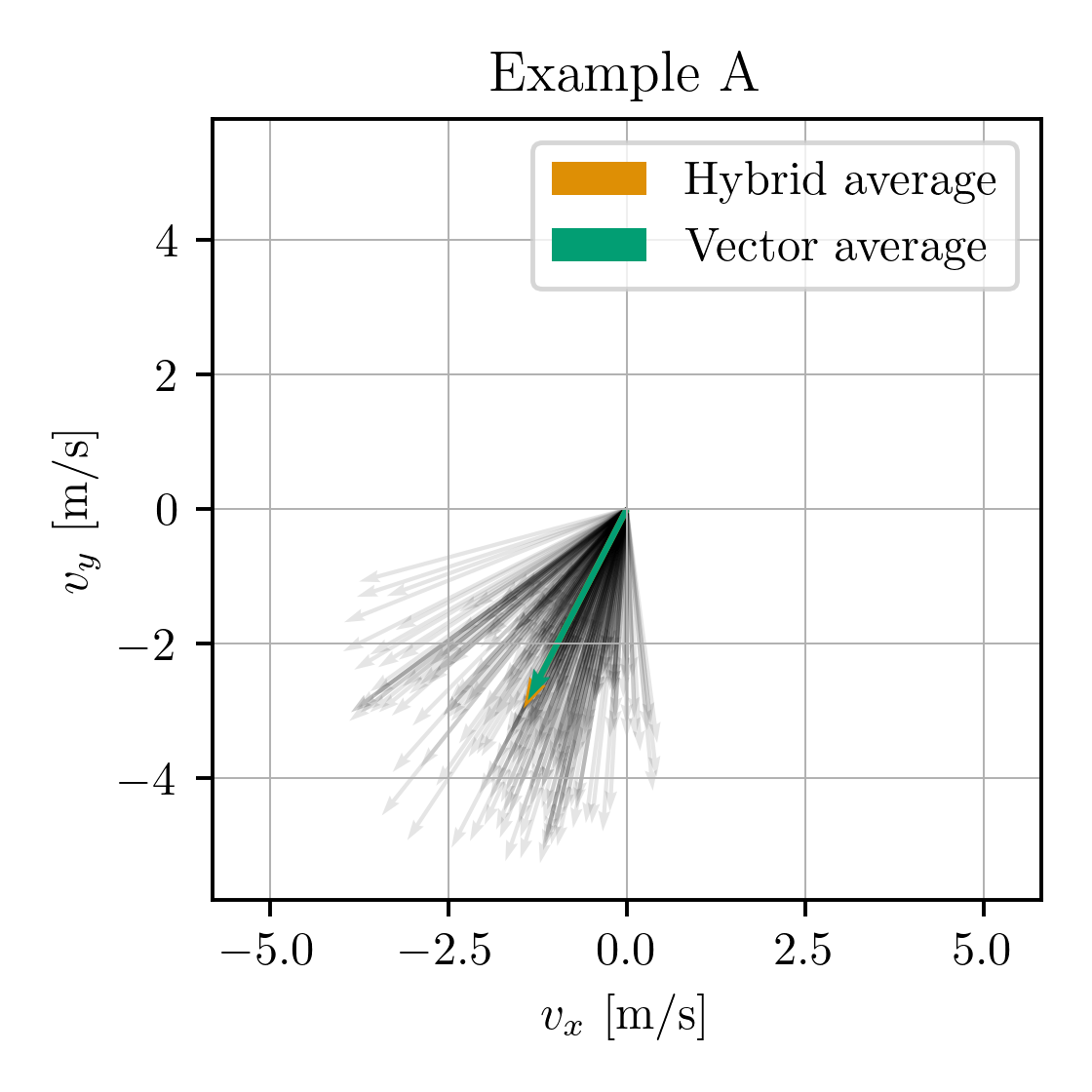}
		\includegraphics[width=0.49\textwidth]{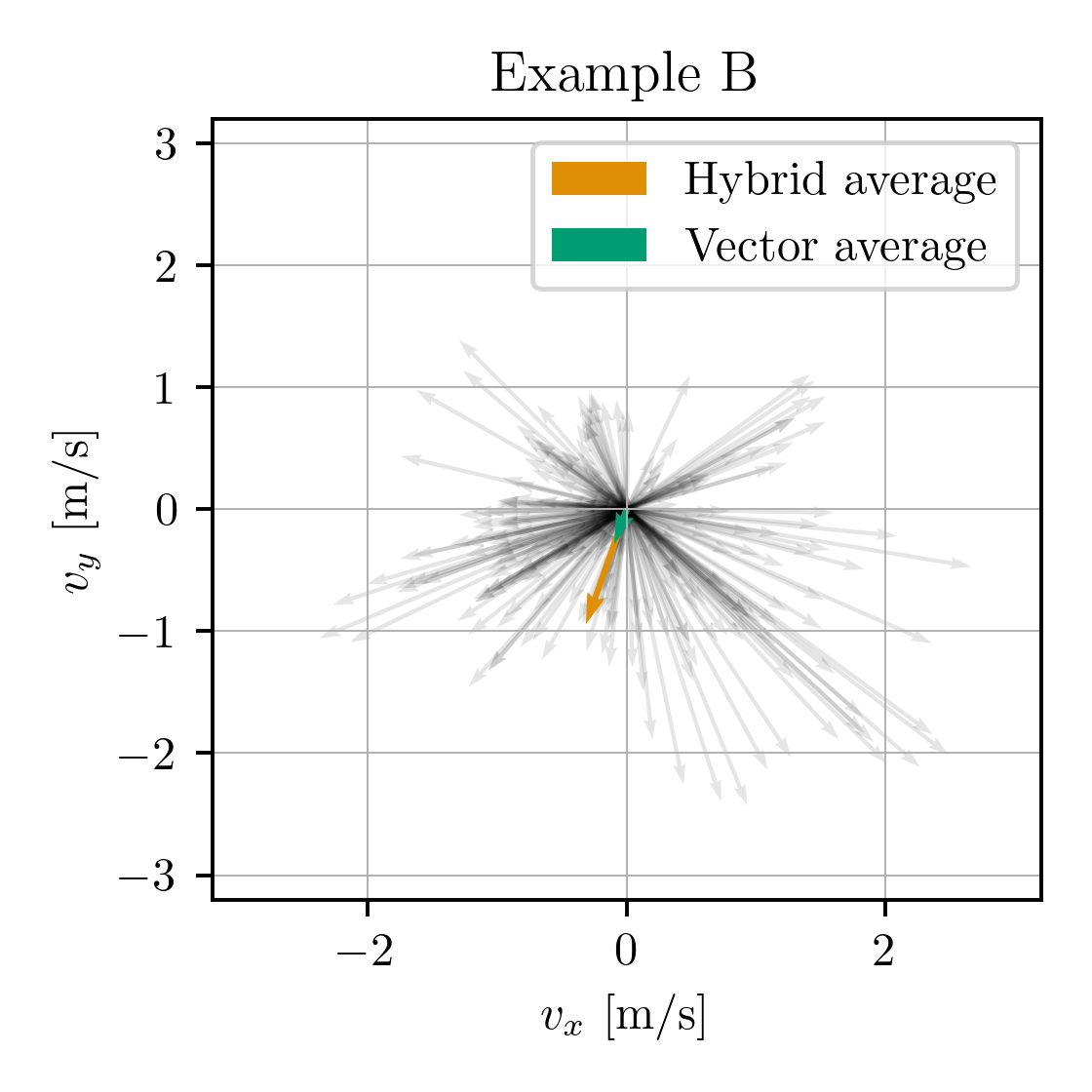}
	\end{center}
	\caption{Vector plot of \SI{1}{\s} wind measurements and average values within one \SI{5}{min} window for example A with higher wind speeds and relatively constant wind direction (left) and example B with lower wind speed and significant variations in wind direction (right).}
	\label{fig:wind_quiver}
\end{figure}

\begin{figure}[H]
	\begin{center}
		\includegraphics[width=0.49\textwidth]{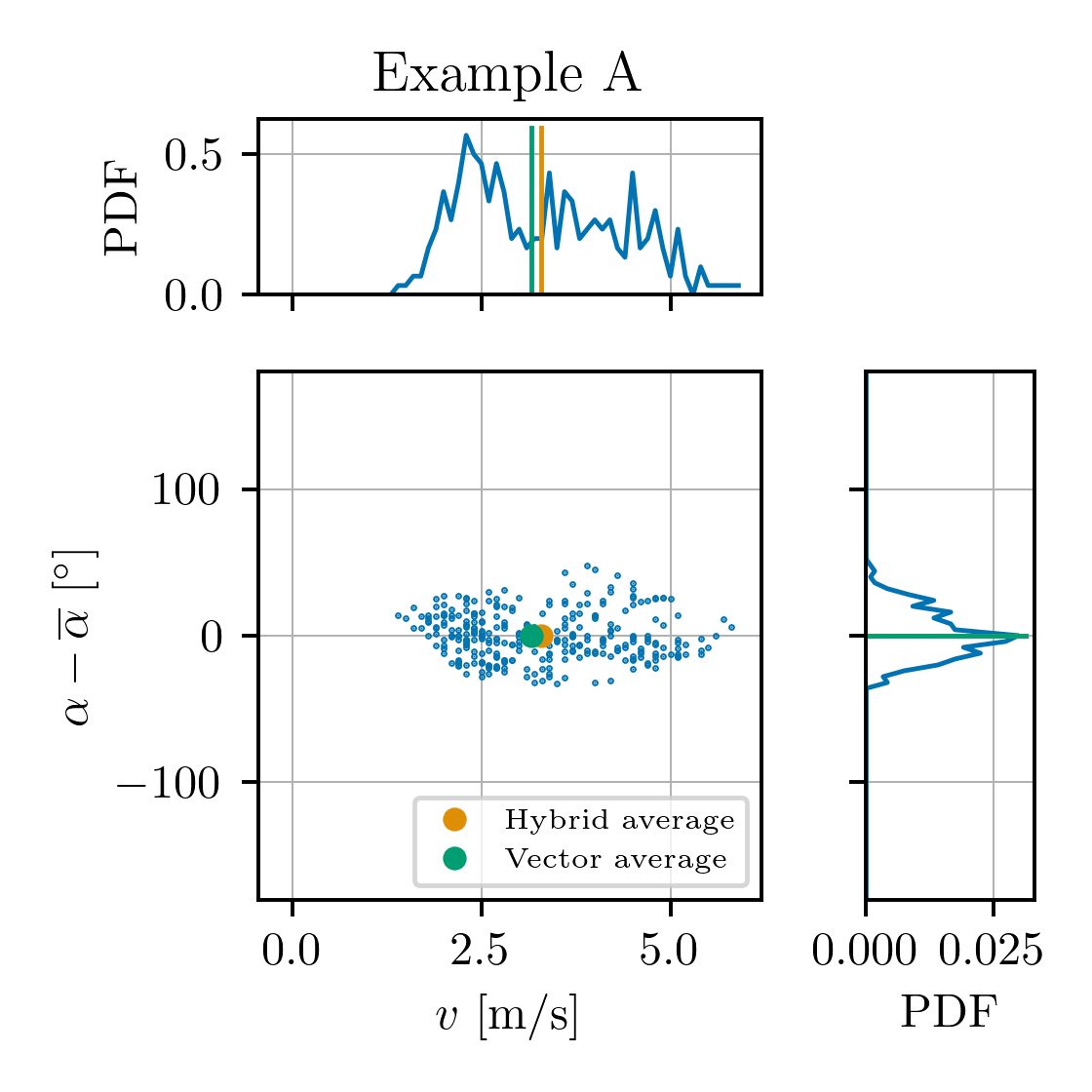}
		\includegraphics[width=0.49\textwidth]{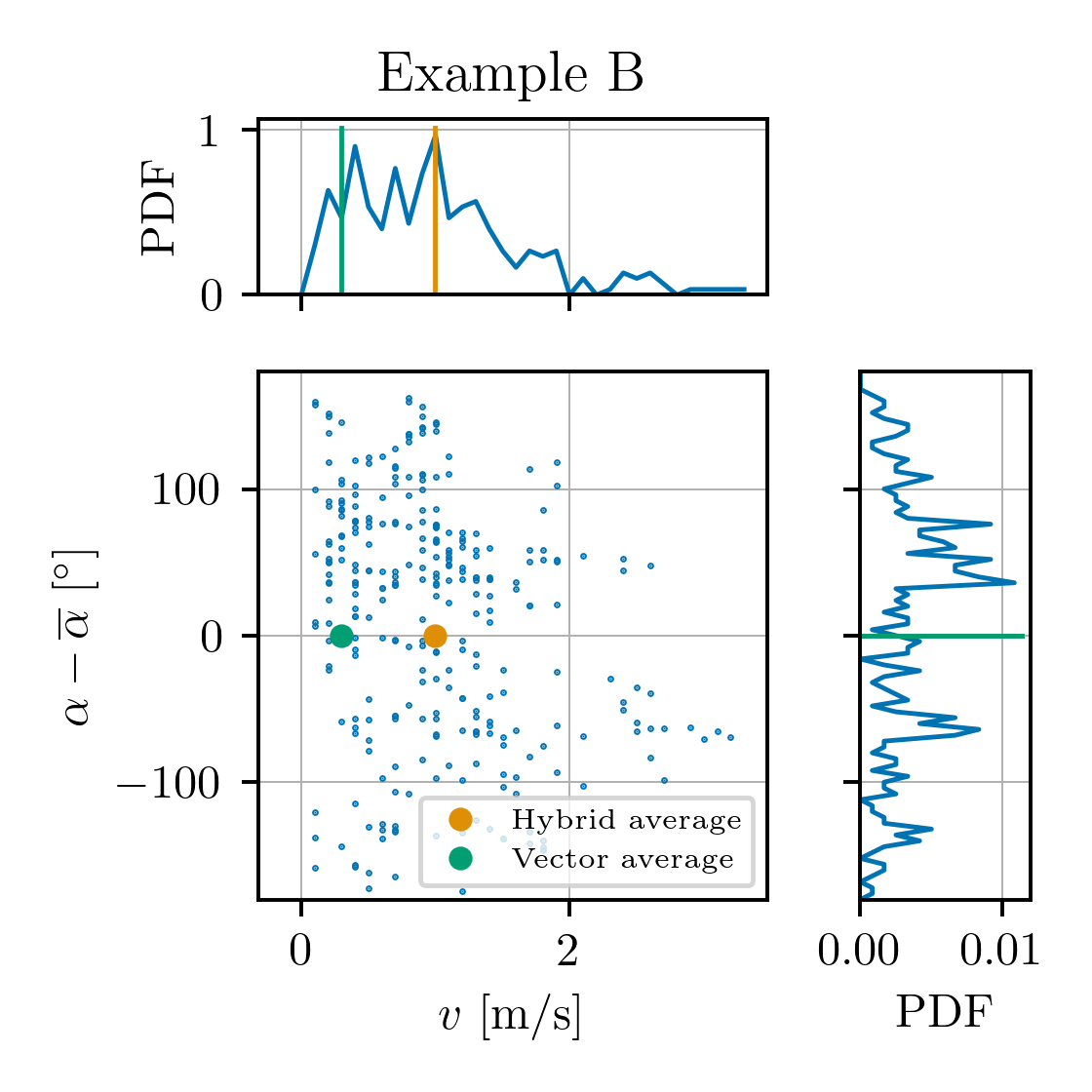}
	\end{center}
	\caption{Scatter plots with speed and direction PDF for example A with higher (left) and example B with lower wind speeds (right).}
	\label{fig:wind_vAndAlpha}
\end{figure}

Let us take a look at both averaging methods in the context of the observed data. Figure~\ref{fig:wind_quiver} presents two examples of wind measurements taken over \SI{5}{\minute} windows. Example A (April 10th, from 14:50 to 14:55) (left) corresponds to above-average wind speeds, approximately \SI{5}{\ms}, where the wind direction remains relatively consistent throughout the window. In contrast, wind in example B (April 4th, from 13:50 to 13:55) (right) has lower speeds, around \SI{1}{\ms}, and fluctuates in direction significantly, covering nearly the entire range of possible directions. We see that in example A, both averaging methods yield similar results, while in example B, there is a considerable difference between the two average wind speeds, which is demonstrated in a more quantitative manner in Figure~\ref{fig:wind_vAndAlpha}.

In addition to the type of averaging, the length of the window itself plays a crucial role, as highlighted in previous studies~\cite{hosek_effect_2011}. As discussed, the lower bound of averaging windows is typically constrained by the system limitations. In this paper, we focus on \SI{5}{\minute} window since this is the default output rate of sensors at the observed location, and we will include a short analysis of the effect of the window length on wind variability and ampacity differences in Section~\ref{Appendix}.

\subsection{Wind variability within the averaging window}
\label{sec:wind-variability-within-the-averaging-window}

In the previous section we established the ground for introducing the concept of the wind variability~\cite{ward_time-averaged_2023, gonzalez-cagigal_influence_2022}. In general, wind variability is discussed across different temporal scales, ranging from minutes to seasons. In the context of this paper, wind variability refers specifically to fluctuations in wind speed and direction over one window, i.e., over short time scales.

There are several ways to characterise wind variability using different metrics~\cite{lee2018assessing}. The most common is the standard deviation of wind speed values over the averaging window. The coefficient of variation is another common measure that expresses wind speed variability as a percentage of the mean wind speed -- essentially a normalised standard deviation. Similarly, turbulence intensity is defined as the ratio of the standard deviation of wind speed to its mean value~\cite{cigre, raichle2009wind}. Additionally, a robust coefficient of variation normalises the median absolute deviation with the median value.

\begin{figure}[H]
	\begin{center}
		\includegraphics[width=0.49\textwidth]{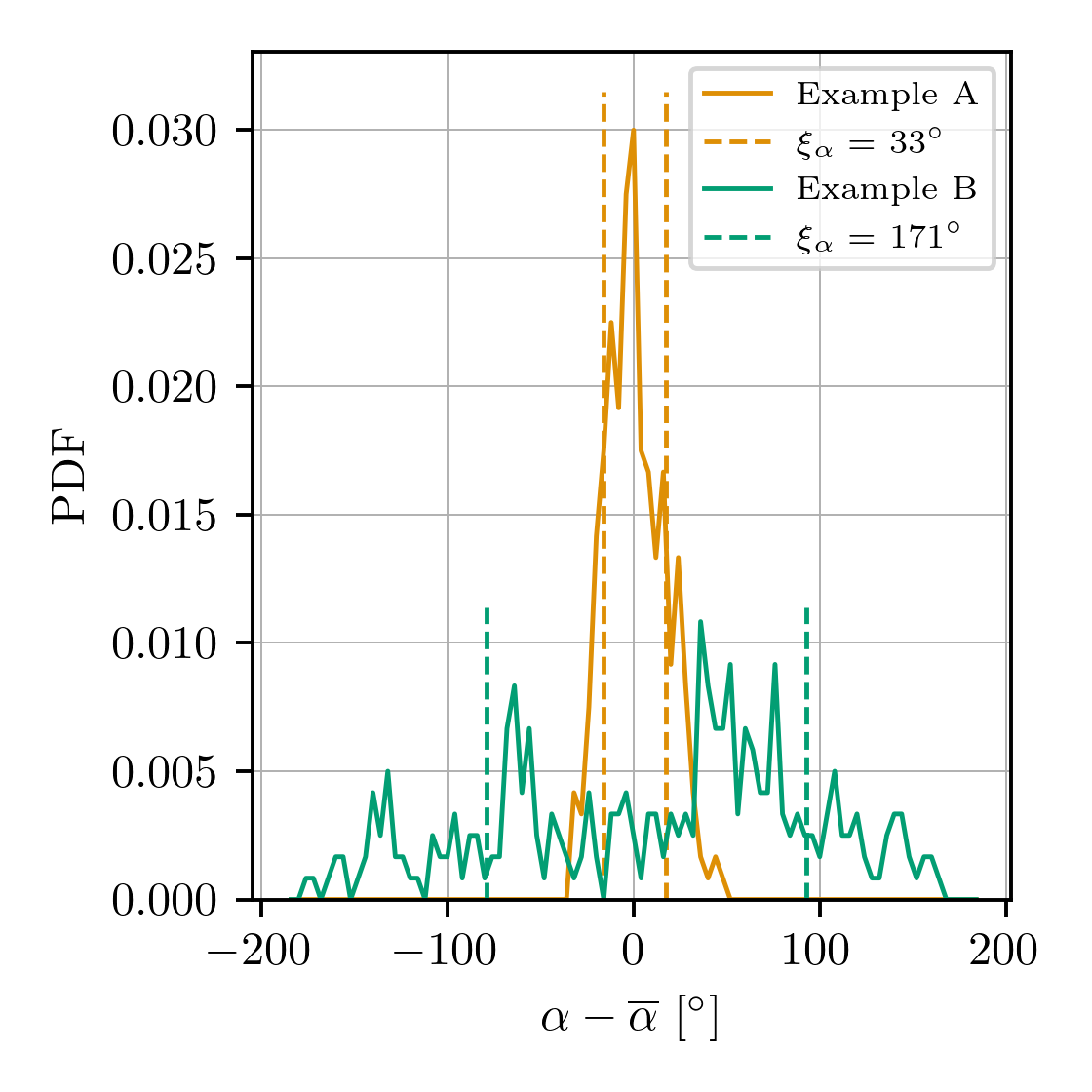}
		\includegraphics[width=0.49\textwidth]{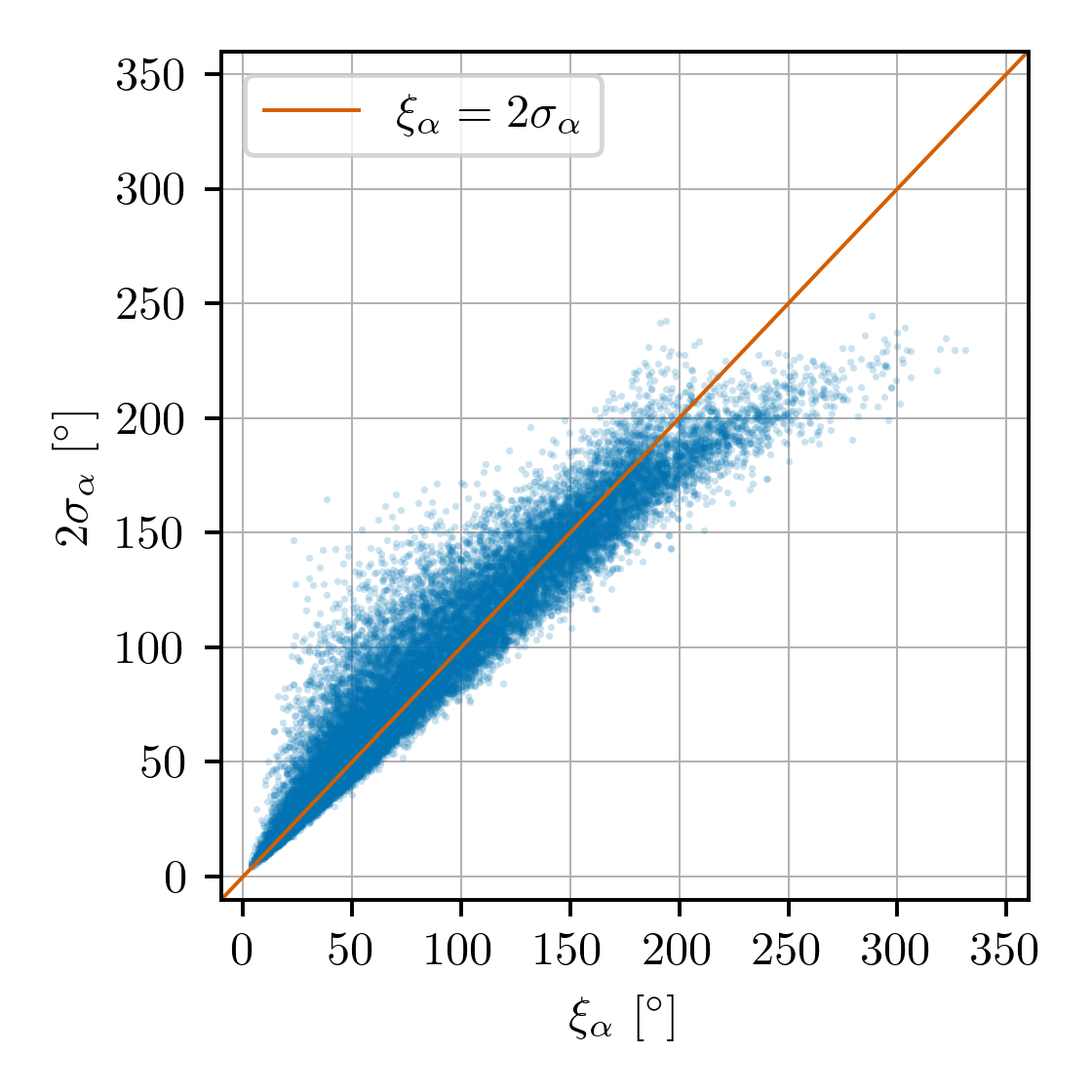}		
	\end{center}
	\caption{Left: Wind direction PDFs for examples A and B along with the upper and lower limits of variability metric. Note that the calculation of the average angle is weighted by wind speed, so the average angle might not seem intuitive, especially for example B. Right: Comparison of variability metric $\xi$ and standard deviation $\sigma$.}
	\label{fig:analysisWindow_spreadDef}
\end{figure}

While most of the above metrics focus on wind speed, in the presented research, attention will be given to both wind speed and direction. We define a variability metric $\xi$ as
\begin{equation}
	\xi_X = X_{p, hi}-X_{p, lo}
	\label{eq:spread}
\end{equation}
where $X$ is either wind speed or direction, and $X_{p, hi}$ and $X_{p, lo}$ are the upper and lower limits for which cumulative distribution function CDF$(X)$ equals $\frac{p}{2}$ and $1-\frac{p}{2}$, respectively. 
For the variability in speed, $\xi_v$, the quantity $X$ is simply the measured speed, while for the direction, $X$ is the direction subtracted by its average in each window and observed on the interval from \ang{-180} to \ang{180} to account for the scale discontinuity, $\xi_{\alpha} \coloneq \xi_{\alpha-\overline{\alpha}}$. To select the optimal value of confidence level $p$, we performed preliminary testing. We found the results to be qualitatively equivalent for a wide range of values, including $p=0.68$, which we decided to use for further experiments. The latter value is special since it causes the proposed metric $\xi$ to become equivalent to standard deviation if $X$ is normally distributed. I.e. $\xi = 2\sigma$, where
\begin{equation}
	\sigma_x = \sqrt{\frac{\sum_{i=1}^n (x_i-\overline{x})^2}{n}}
	\label{eq:stdDev}
\end{equation}
with $x_i$ representing the $i$-th measurement, $\overline{x}$ is the mean of all the measurements, and $n$ is the number of measurements. Generally however, the speed and direction are not normally distributed. Especially with the direction, we have observed a plethora of different shapes, including multi-modal, which makes defining a variability measure challenging and poses an opportunity for further studies. In a multi-modal case, we believe the variability metric $\xi_X$ holds more information than the standard deviation, as it takes the lower and upper limits of $X$ into account, which hold information about the asymmetry of the distribution. Figure~\ref{fig:analysisWindow_spreadDef}~(left) shows the wind direction distributions for examples A and B, along with the appropriate upper and lower limits of variability metric. Figure~\ref{fig:analysisWindow_spreadDef}~(right) shows a comparison of the variability metric and standard deviation. We see that the metrics are correlated, however they are not identical. The spread we get is the consequence of the fact that two distributions with radically different shapes may share the same $\xi_{\alpha}$.

\begin{figure}[H]
	\begin{center}
		\includegraphics[width=0.49\textwidth]{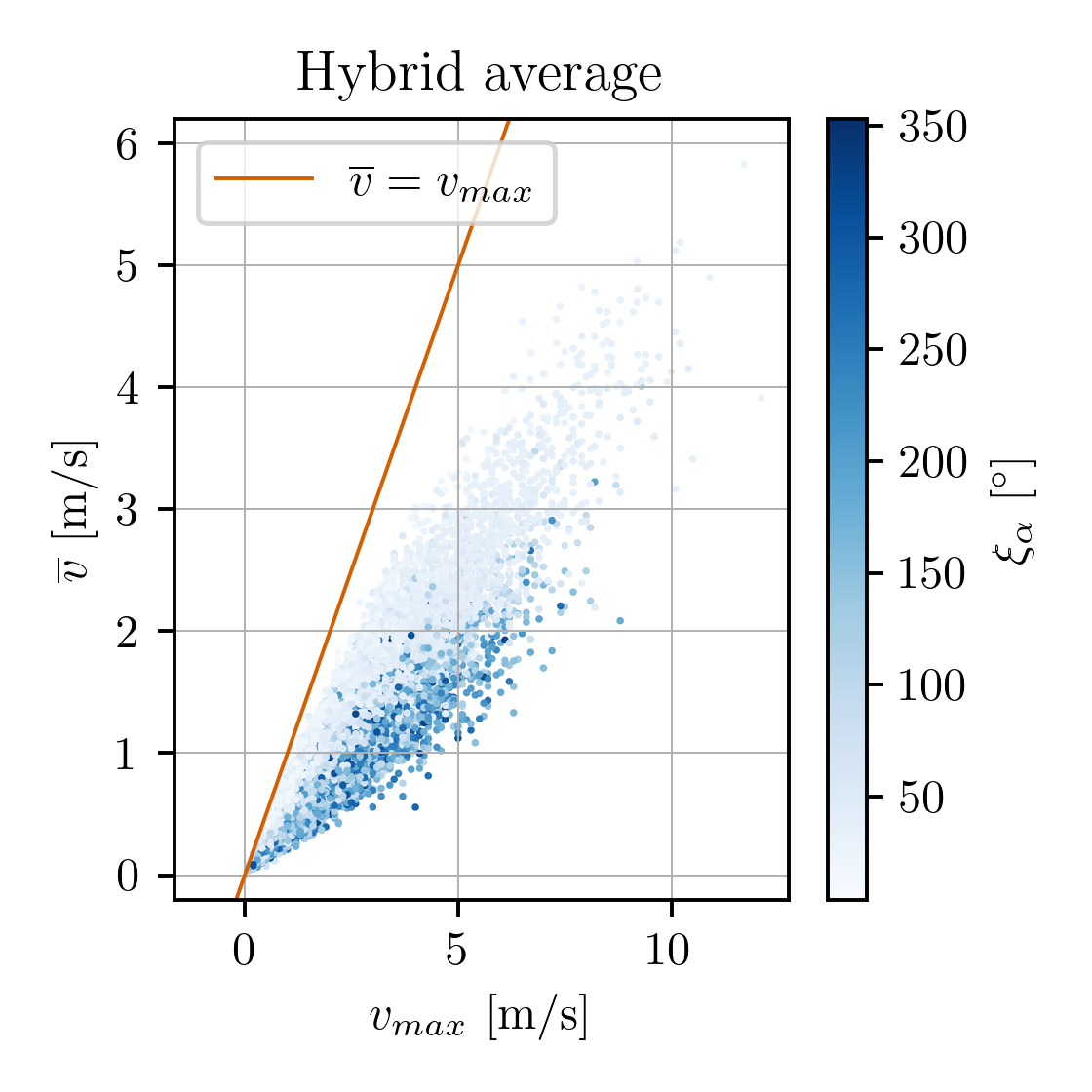}
		\includegraphics[width=0.49\textwidth]{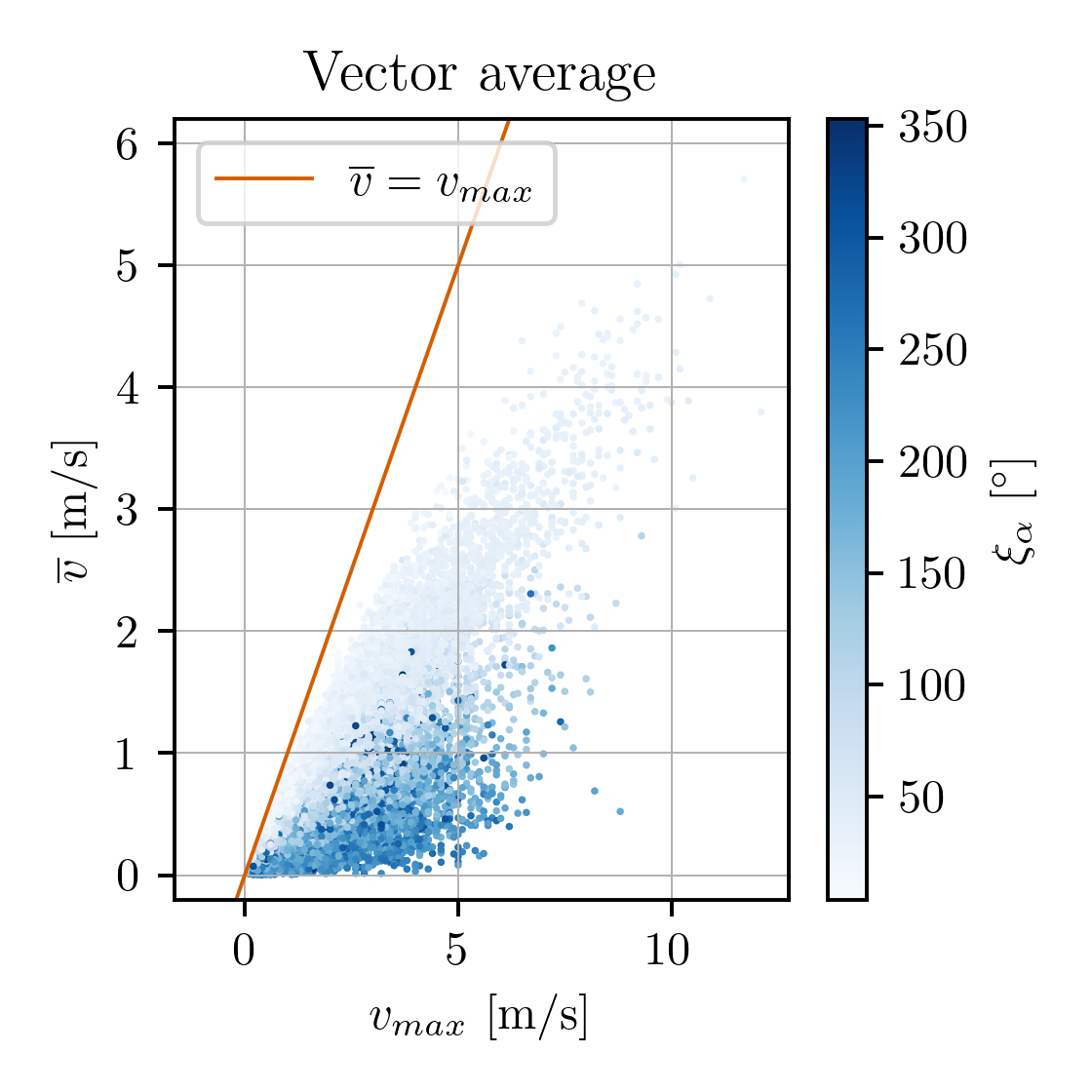}
		\includegraphics[width=0.49\textwidth]{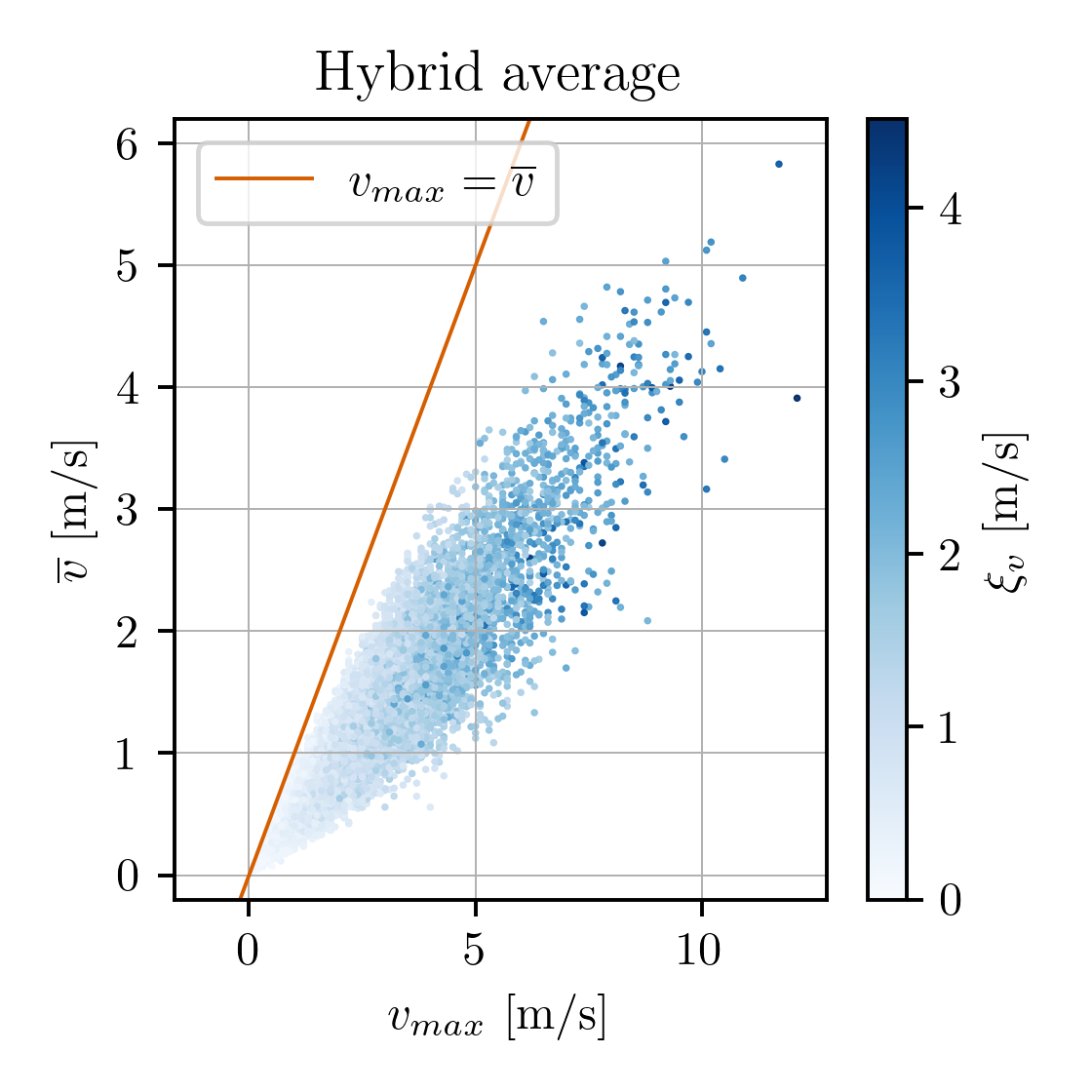}
		\includegraphics[width=0.49\textwidth]{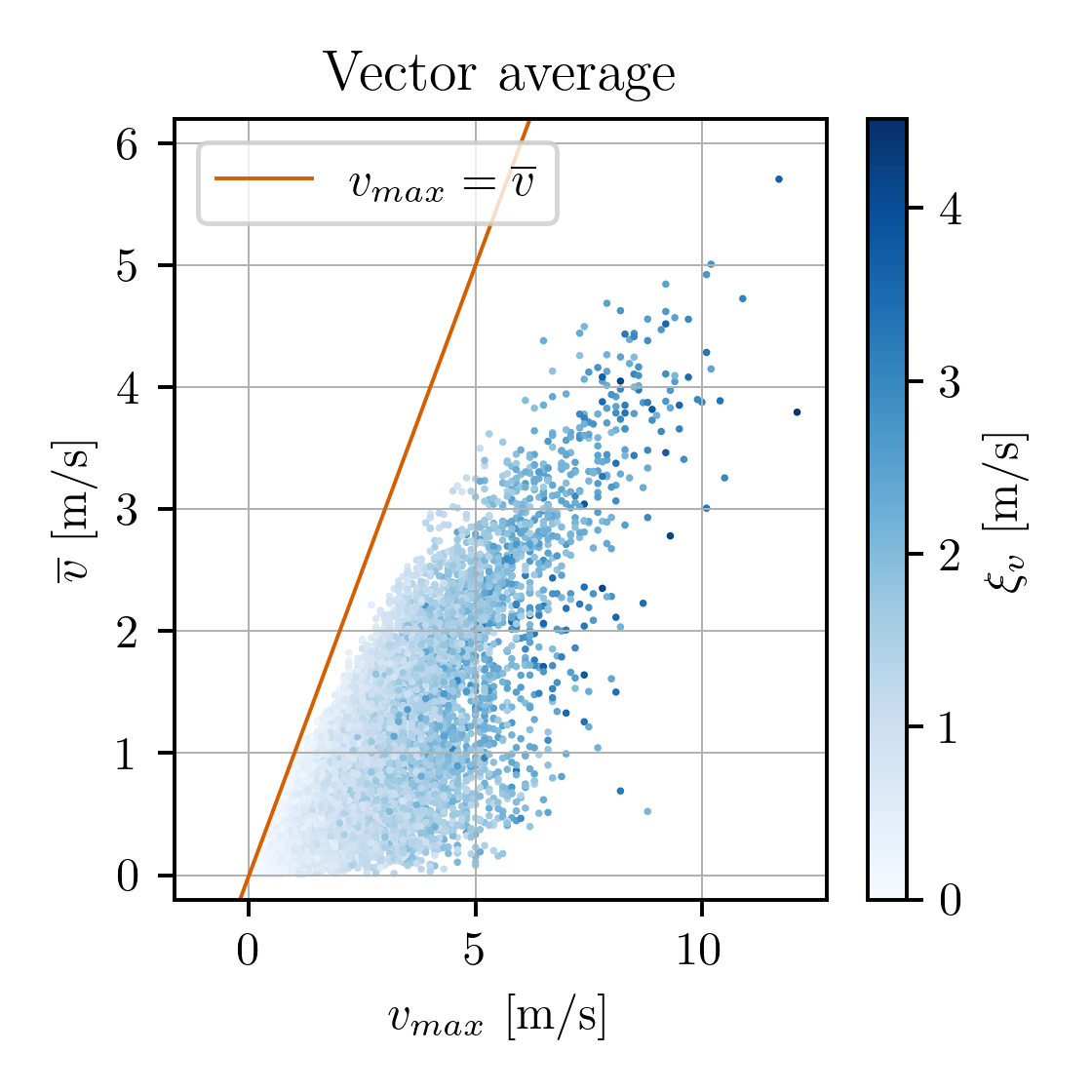}		
	\end{center}
	\caption{Correlation of $v_{max}$ and $\overline{v}$ within all windows in the context of $\xi_{\alpha}$ and $\xi_v$.}
	\label{fig:wind_averagingSpeeds}
\end{figure}

We can now look at the two averaging methods and their effect on the wind speed in the context of the variability metric. The insights will be crucial for understanding why each averaging method impacts ampacity differently. Figure~\ref{fig:wind_averagingSpeeds} compares the average speed for each \SI{5}{\minute} window with the maximum measured speed within that window, with the $\xi_{\alpha}$ and $\xi_v$ represented with colour. In case of zero variability in both speed and direction, the average and max speed would be equal. Let us first take a look at the variability in wind direction (top row). We note two things, which we have already observed in the two selected examples in Figure~\ref{fig:wind_quiver}. First, for both averaging methods, the amplitude of the average speed falls with the increase in $\xi_{\alpha}$, and second, the effect is much more pronounced in vector averaging (right), where the scatter plot spreads towards low $\overline{v}$ values, and we get cases where we have a considerable maximum speed, but the average speed is almost zero, which effectively gives us no wind. In this case, the average speed is not representative of the speeds within the window.

Now to the variability in wind speed (bottom row). We can see the combination of two trends. First, and this is quite pronounced, $\xi_v$ increases with $v_{max}$. At the same time, at each $v_{max}$, the point with higher $\xi_v$ will have a lower $\overline{v}$, which is similar to what we have seen above, though less pronounced. Let us look at the top envelope of the scatter plot. We have seen above that this is where we get minimal variability in wind direction. The distance between the envelope and the $\overline{v}=v_{max}$ (the limit where $\xi_v=0$ and $\xi_{\alpha}=0$) tells us something about the effect of the variability in speed. We can see that the distance is growing with $v_{max}$, which is in line with the observation that $\xi_v$ is also growing with $v_{max}$.

\begin{figure}[H]
	\begin{center}
		\includegraphics[width=0.49\textwidth]{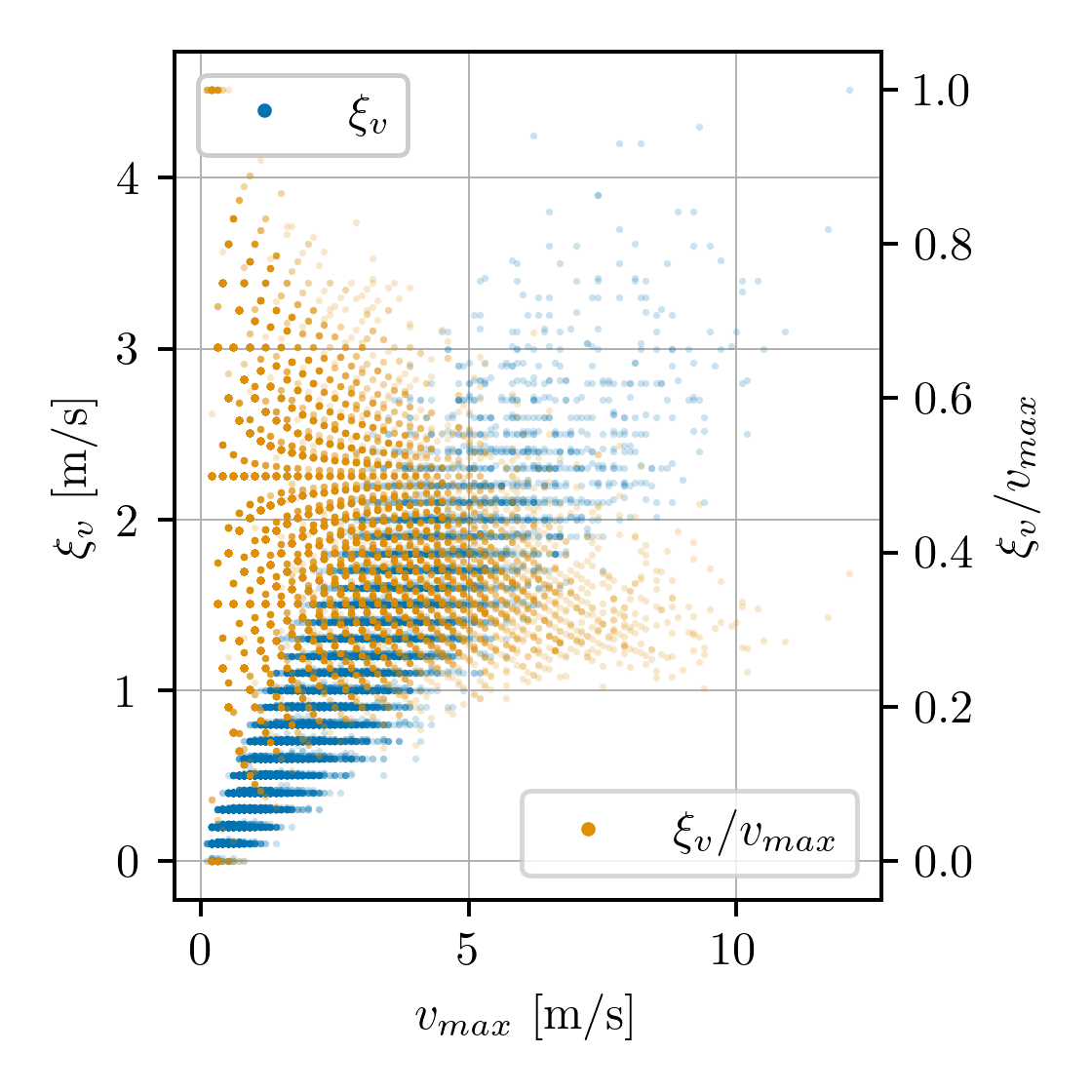}
		\includegraphics[width=0.49\textwidth]{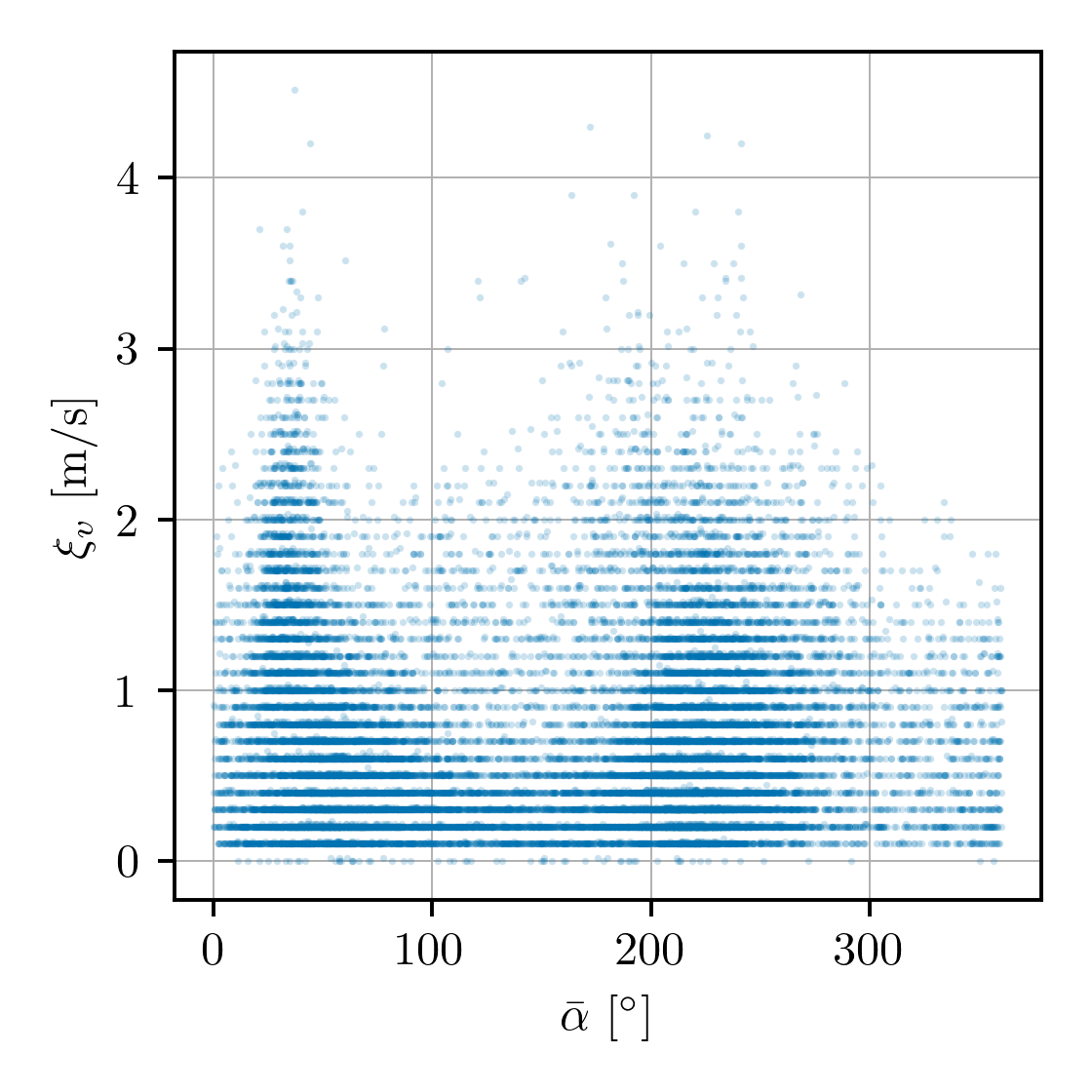}
		\includegraphics[width=0.49\textwidth]{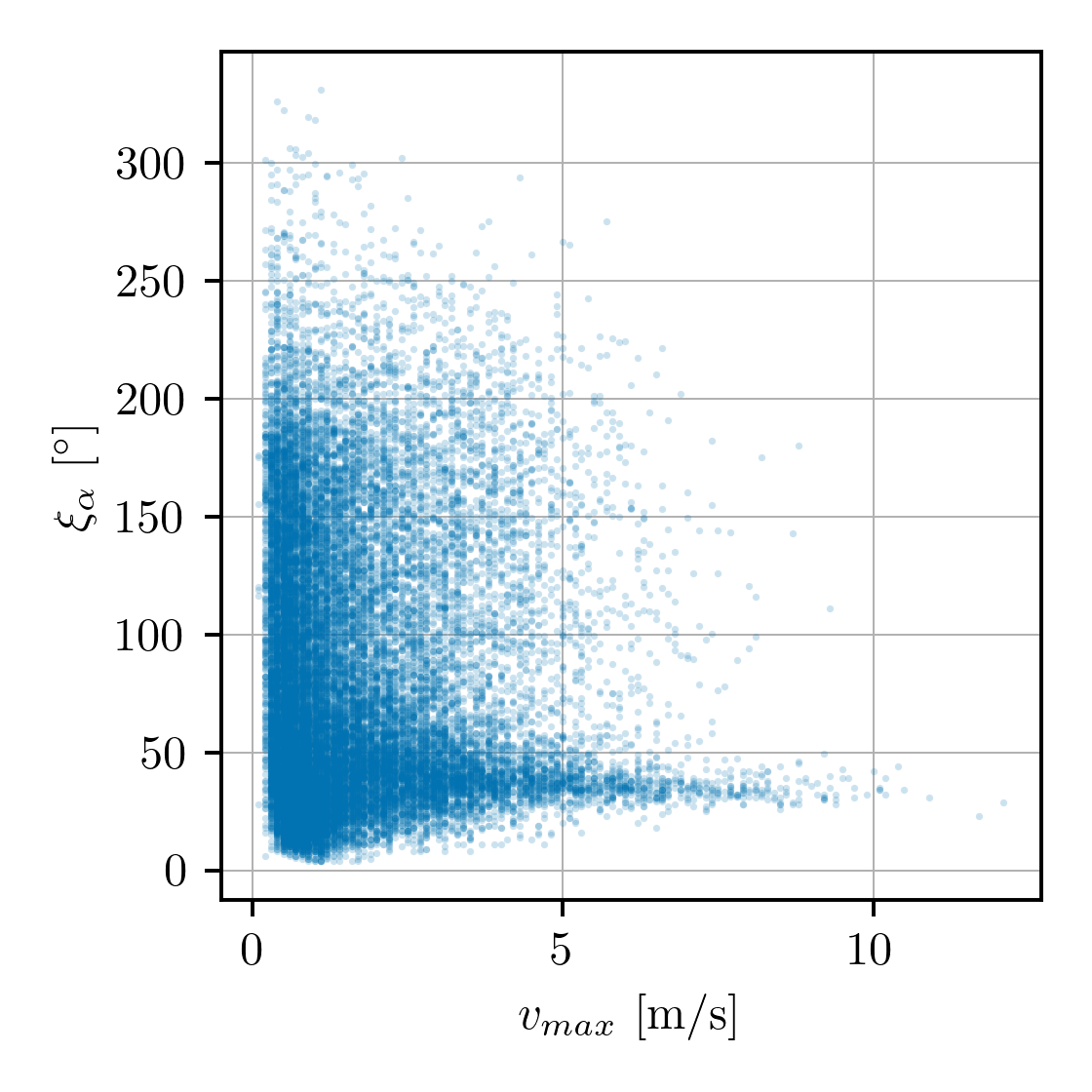}
		\includegraphics[width=0.49\textwidth]{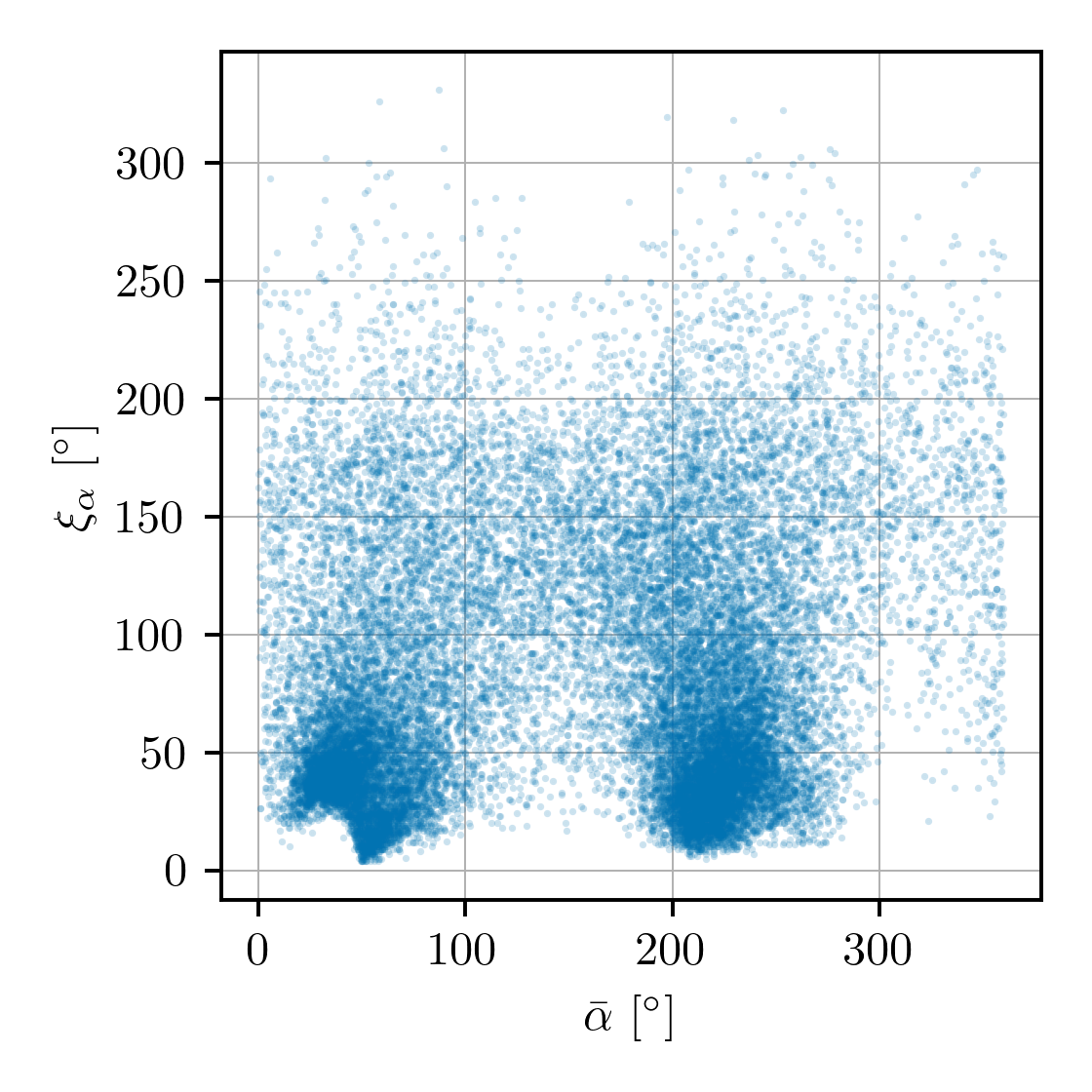}
	\end{center}
	\caption{The dependence of variability in wind speed and direction on the speed and direction.}
	\label{fig:wind_spreads}
\end{figure}

Figure~\ref{fig:wind_spreads} shows the dependence of variability $\xi$ over all of the 5-minute windows on the wind speed and direction. Note that the maximum wind speed is plotted on the graph (as opposed to the average speed), so the observations are independent of the averaging method. The variability in wind speed $\xi_v$ shows a linear correlation with speed (top left), and if we take a look at the relative variability, $\xi_v/v_{max}$, it appears to reach a plateau value of around 0.3 for larger wind speeds for the observed location. This is consistent with findings in~\cite{arenasLopez2020stochastic}, which focused on a location in Mexico, so this observation transfers across more than one location. On the other hand, the patterns in the $\xi_v$ and average direction plot (top right) are likely the consequence of the speed-direction relation for the observed location with the two prominent peaks matching the location of peaks in Figure~\ref{fig:wind_vAndAlphaSource}. The $\overline{\alpha}$ vs $\xi_{\alpha}$ plot (bottom right) seems to have an envelope, especially at very low variabilities, which seem to be observed only in conjunction with one of the prevailing wind direction observed in Figure~\ref{fig:wind_vAndAlphaSource}, and are therefore also location-specific.

The variability in wind direction $\xi_{\alpha}$ in relation to wind speed in Figure~\ref{fig:wind_spreads} (bottom left) has several points of interest. First, it has a prominent peak (in the horizontal direction) at around \ang{35}, followed by a plateau and a decrease for larger wind variabilities in direction. Note that the position of the decrease is influenced by the selection of parameter $p$. This peak-and-decrease shape is in agreement with the two characteristic examples with high and low wind speeds from Figure~\ref{fig:wind_quiver}, where example A with higher wind speed had better-defined wind direction, and in example B with lower wind speed, the wind direction was fluctuating considerably. It is also in agreement with the general notion that wind does not have a predominant direction at low speeds~\cite{cigre}. What is more interesting is the bottom section of the graph, where we find that as the wind speed increases, the minimal observed variability in wind direction also increases, until it plateaus at around \ang{35}. In other words, at higher wind speeds, there is an inherent variability in wind direction. This could be the effect of wind turbulence.

\section{Effect of wind variability on DTR} \label{sec:dtr}

Now, let us take a look at the mechanism with which wind variability affects the DTR results. In order to do that, we will briefly look at the DTR model equations. As the subject of the experiments, we take an overhead power line which comprises a steel core and aluminium conductor. In principle, there is Joule heating in the conductor, heat transfer in both the conductor and core, and a set of boundary conditions: symmetry in the centre, temperature and heat flux continuity on the core-conductor interface, and heat exchange with the surroundings at the conductor skin. The calculations of DTR in this paper are performed according to CIGRE guide~\cite{cigre}, which simplifies the radial dependency part and solves the following heat equation:
\begin{equation}
    (\rho_{St}c_{p, St}S_{St} + \rho_{Al}c_{p, Al}S_{Al})\frac{\partial T_{av}}{\partial t}  = Q_J + Q_s + Q_r + Q_c\quad \left[ \unit{\watt\per\metre}\right]
    \label{eq:cigre_heat}
\end{equation}
where $\rho$ are the densities of the steel core and aluminium conductor, $c_p$ are the appropriate specific heat capacities and $S$ are the corresponding cross-section areas. $Q$ are the heat contributions, namely Joule heating, solar heating, radiative cooling and convective cooling and $T_{av}$ is the average temperature of the line. CIGRE guide assumes that the latter is an arithmetic average of the core and skin temperature, $T_{av}=(T_c+T_s)/2$ and that heat generation in the line is uniform, simplifying the radial temperature dependency of a cylindrical conductor into
\begin{equation}
	T_c - T_s = \frac{Q_J}{2\pi\lambda}\left[ \frac{1}{2}-\frac{r_1^2}{r_2^2-r_1^2}\ln\left( \frac{r_2}{r_1}\right) \right] \quad \left[ \unit{\degreeCelsius}\right]
\end{equation}
where $r_1$ and $r_2$ are the radii of the steel core and the overall diameter of the conductor, respectively, and $\lambda$ is the effective radial thermal conductivity. Let us take a look at the individual heat contributions: the Joule heating for alternating current is given by
\begin{equation}
    Q_J = k_s I^2 R_{dc}\quad \left[ \unit{\watt\per\metre}\right]
\end{equation}
with $I$ being electric current, $k_s$ skin effect factor, which is a scalar constant, characteristic of the line, and $R_{dc}$ direct current resistance, which depends on the temperature. CIGRE gives additional equations for its calculation. Next, we have solar heating
\begin{equation}
    Q_s = 2\alpha I_s r_2\quad \left[ \unit{\watt\per\metre}\right]
\end{equation}
with $\alpha$ being conductor surface absorptivity, characteristic of the line, and $I_s$ is the global radiation intensity, which can be either measured or estimated from geographical location. Now to the cooling contributions, we first have the radiation cooling
\begin{equation}
    Q_r = -2\pi r_2 \sigma_B \epsilon\left[ T_{s}^4-T_{a}^4\right] \quad \left[ \unit{\watt\per\metre}\right]
\end{equation}
with $\sigma_B$ the Stefan-Boltzmann constant, $\epsilon$ is conductor surface emissivity, characteristic of the line, and $T_a$ is the ambient (air) temperature, which is usually measured. The last cooling contribution from \ref{eq:cigre_heat}, is convection
\begin{equation}
    Q_c = -\pi \lambda_f (T_s-T_a)\mathit{Nu}\quad \left[ \unit{\watt\per\metre}\right]
    \label{eq:convection}
\end{equation}
where $\lambda_f$ is thermal conductivity of air at film temperature, $T_f = (T_s+T_a)/2$, given in CIGRE. $\mathit{Nu}$ is Nusselt number.

In the case of zero wind, convective cooling is present in the form of natural convection, where the air, heated by the conductor, rises and creates a flow around the conductor. The Nusselt number for natural convection is given by
\begin{equation}
    \nbr{Nu}_{natural\ convection} = A(\nbr{Gr}\nbr{Pr})^m
\end{equation}
where $\nbr{Gr}$ is Grashof number and $\nbr{Pr}$ is Prandtl number and $A$ and $m$ are scalar parameters that are given in tables for different values of $\nbr{Gr}\nbr{Pr}$. 

For high wind speed, the Nusselt number for the perpendicular flow is given by
\begin{equation}
    \nbr{Nu_{90}} = B\nbr{Re}^n
    \label{dtr:nu90}
\end{equation}
with Reynolds number $\nbr{Re} = 2r_2v/\nu_f$; $v$ is wind speed, and $\nu_f$ is kinematic viscosity of air at film temperature. Parameters $B$ and $n$ are given in tables and depend on both $\nbr{Re}$ and the roughness of the conductor surface. Historically, the correlation between Nusselt and Reynolds number was proposed by McAdams, who studied the heat transfer by forced convection in smooth circular cylinders~\cite{morgan_rating_1967}. It was followed by experiments with roughened cylinders of different configurations~\cite{morgan_heat_1973} and the perpendicular flow. The case of interest for DTR were the stranded cylinders of the shape of the bare stranded conductors, and several authors conducted wind tunnel tests and field experiments as well as proposed the correlation functions~\cite{cigre,morgan_rating_1967,waghorne_current_1951,fand_continuous_1972}. CIGRE relies heavily on work by Morgan -- he performed several studies in the sixties and seventies using wind tunnels~\cite{morgan_rating_1967} and proposed the full DTR model in 1982~\cite{morgan_thermal_1982}. 
The wind tunnel tests were performed with very low turbulence (i.e. variation in wind speed)~\cite{morgan_heat_1973}, and with flow velocity uniform across the working section. CIGRE guide notes that the relations can be considered as the local performance of the conductor in constant laminar wind, and that the Nusselt number increases with the intensity and scale of the turbulence, but it does not provide any relations, and states that the turbulence assessment is very complicated for real-life installations.

In the context of this study and the high temporal resolution of the observed data, we will use these relations for short time scales, in an effort to make the first exploratory step to high-resolution DTR. However, further studies are needed to explore the validity of the relations over different (shorter) time scales, and variable wind. We see this as a potential for future work.

With the relation between the Nusselt number and Reynolds number established for perpendicular flow, experiments were repeated for various relative angles of the flow. CIGRE guide proposes the following correction for the wind direction relative to the conductor axis $\alpha_{rel}$, in relation to the \nbr{Nu_{90}} value
\begin{equation}
\begin{split}
    \nbr{Nu} &= \nbr{Nu_{90}}(0.42 + 0.68(\sin(\alpha_{rel}))^{1.08})\mathrm{,\ }\alpha_{rel}\leq 24 \unit{\degree} \\
    \nbr{Nu} &= \nbr{Nu_{90}}(0.42 + 0.58(\sin(\alpha_{rel}))^{0.90})\mathrm{,\ }\alpha_{rel}> 24 \unit{\degree}.
\end{split}
\label{eq:Cigre_NuAngle}
\end{equation}
In practice, the transition between the natural and forced convection occurs at around \SI{0.5}{\ms}, and there are several models proposed for the transition. However, CIGRE recommends simplifying the selection of regime at low wind speeds by calculating both forced and natural convection, then using the higher of the two values. 

Finally, heat equation~\ref{eq:cigre_heat} can be rearranged to compute ampacity, i.e., the maximum current in the line that will not cause the line to exceed its thermal rating at the given weather conditions
\begin{equation}
    I_{th} = \sqrt{-\frac{Q_c(T_{max}) + Q_r(T_{max}) + Q_s(T_{max})}{k_sR_{dc}(T_{max})}}\quad [\unit{\ampere}]
\end{equation}
where $T_{max}$ is the maximum allowed conductor temperature. In Slovenia, $T_{max}=\SI{80}{\degreeCelsius}$, and this value is used in the calculations.

To isolate the effect of wind measurements, which are the core of the presented study, all of the other weather parameters are set to constant values for further analysis. To set realistic conditions that would be of interest to TSOs, solar radiation and temperature are set to match a typical sunny day, with an average temperature for timespan between April (the coldest of the observed months) and August (the warmest of the observed months) for the observed location. I.e. $I_s=\SI{900}{\watt\per\metre}$, and $T_a=\SI{15}{\degreeCelsius}$. The observed line span lies between pylons SM111 and SM112 on Figure~\ref{fig:wind_windrose}, which means the line direction at the point of interest is north-south and the prevailing wind directions are neither parallel nor perpendicular.

In the used heat model, wind speed and direction are featured only in the convective term, in the Nusselt number calculation, so this is the minimal building block on which we can study the wind effect. However, in practice, the quantity discussed most is ampacity, especially as it is of special interest to TSOs, so we will show the results in the context of ampacity too. 

\begin{figure}[H]
	\begin{center}
		\includegraphics[width=0.49\textwidth]{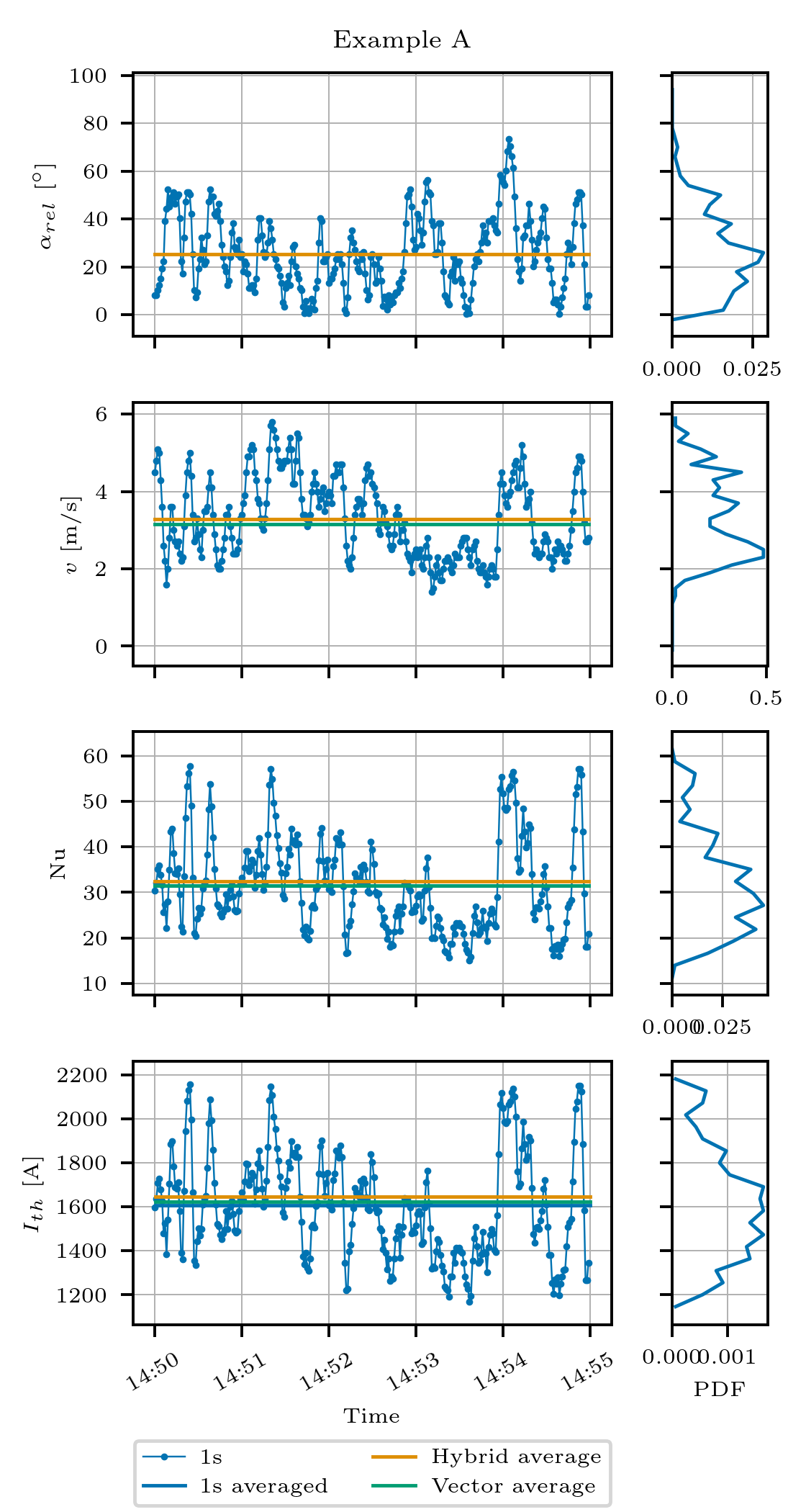}		
		\includegraphics[width=0.49\textwidth]{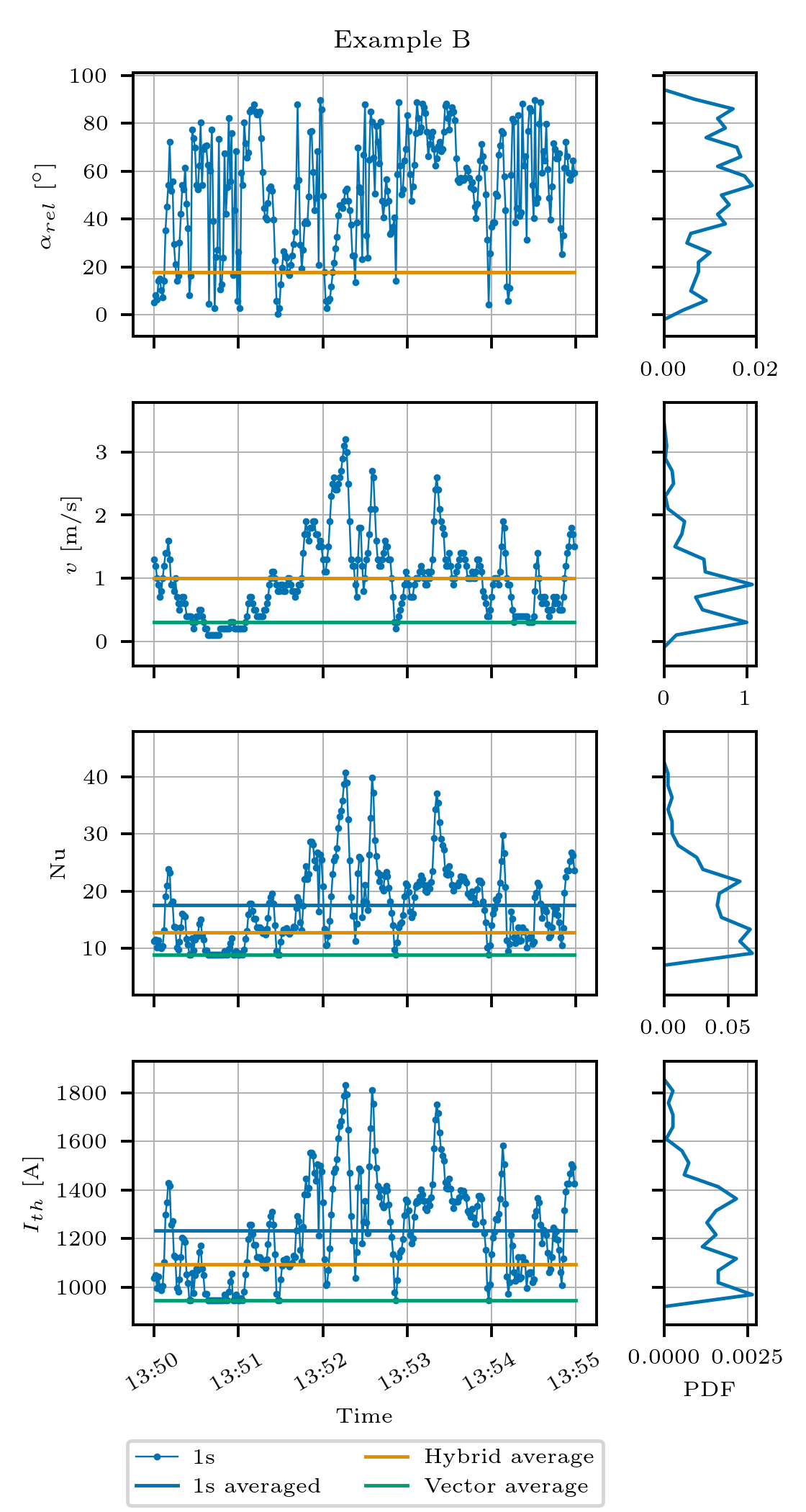}
	\end{center}
	\caption{Wind measurements with relative wind direction and wind speed, Nusselt number and ampacity calculations for an example with higher wind speeds and low $\xi_{\alpha}$ (left) and lower wind speeds and high $\xi_{\alpha}$ for both averaging methods. Note that by definition, the average angle is the same in vector and hybrid averaging used in this study.}
	\label{fig:dtr_timeline}
\end{figure}

With the model and the wind variability metrics defined, we can now take a look at some real-life data. First, let's take a look at two illustrative examples of how the combination of wind variability and averaging method affects the results. Figure~\ref{fig:dtr_timeline} shows the wind speed and direction timelines of examples A and B along with the calculated Nusselt number and ampacity. The high-resolution calculations are performed with \SI{1}{\second} wind measurements, and the average of these calculations, \Nuavg\ and \Iavg will serve as the benchmark for comparison. They will be compared to the Nusselt numbers and ampacities calculated on \SI{5}{\minute} averaged wind measurements, $\nbr{Nu}(\overline{w})$ and $I_{th}(\overline{w})$, for both averaging methods.

In example A, the 3 Nusselt numbers are within \SI{4}{\percent} of each other, with the Nusselt number on hybrid-averaged data above, and Nusselt number on vector-averaged just below the average Nusselt number of \SI{1}{\second} data \Nuavg. The three ampacities are within \SI{45}{\ampere}, which represents around \SI{3}{\percent} of the value, with both ampacities calculated on the averaged data being just above the \Iavg. The small difference in results is expected, as we have seen that with well-defined wind direction and low $\xi_{\alpha}$, the scalar and vector average of the wind speed are close together (also within \SI{4}{\percent} of each other as shown on the same figure), and speed averaged in either way is representative of the whole window. Example A clearly represents a favourable situation for using the averages in DTR calculations. 

On the other hand, in example B, the difference in Nusselt numbers is significant. The hybrid-averaged data Nusselt number is around \SI{27}{\percent} lower than \Nuavg, and the vector-averaged data Nusselt number is almost \SI{50}{\percent} lower than \Nuavg. There are two effects that contribute to the differences. First, vector-averaged speed is significantly lower than the scalar-averaged speed (almost \SI{80}{\percent} lower), which is due to large $\xi_{\alpha}$, and the spread towards lower average speeds we have seen in Figure~\ref{fig:wind_averagingSpeeds}. This explains why the Nusselt number is lower for vector-averaged data than for the hybrid-averaged data. Second, if we look at the relative angle distribution, we see that most values lie above \ang{40}, which is favourable for convective cooling. But the average relative angle is much lower, around \ang{19}. We can understand this if we take a look at Figure~\ref{fig:wind_quiver} (right), where we see that we get significant wind contributions from the east- and westward directions, however, the average angle gives us wind from the north-north-east. With the observed line span in the north-south direction, the averaging results in a lower relative angle which creates a less favourable cooling scenario, and this is true for both averaging methods, as wind direction is averaged with vector averaging in both examples. This is why both Nusselt numbers calculated on averaged data are lower than \Nuavg. Looking at the ampacities, the hybrid-averaged data ampacity is around \SI{140}{\ampere} (\SI{11}{\percent}) lower than \Iavg, and vector-averaged data ampacity is around \SI{290}{\ampere} (\SI{23}{\percent}) lower than \Iavg. Both differences are substantial.

\begin{figure}[H]
	\begin{center}
		\includegraphics[width=\textwidth]{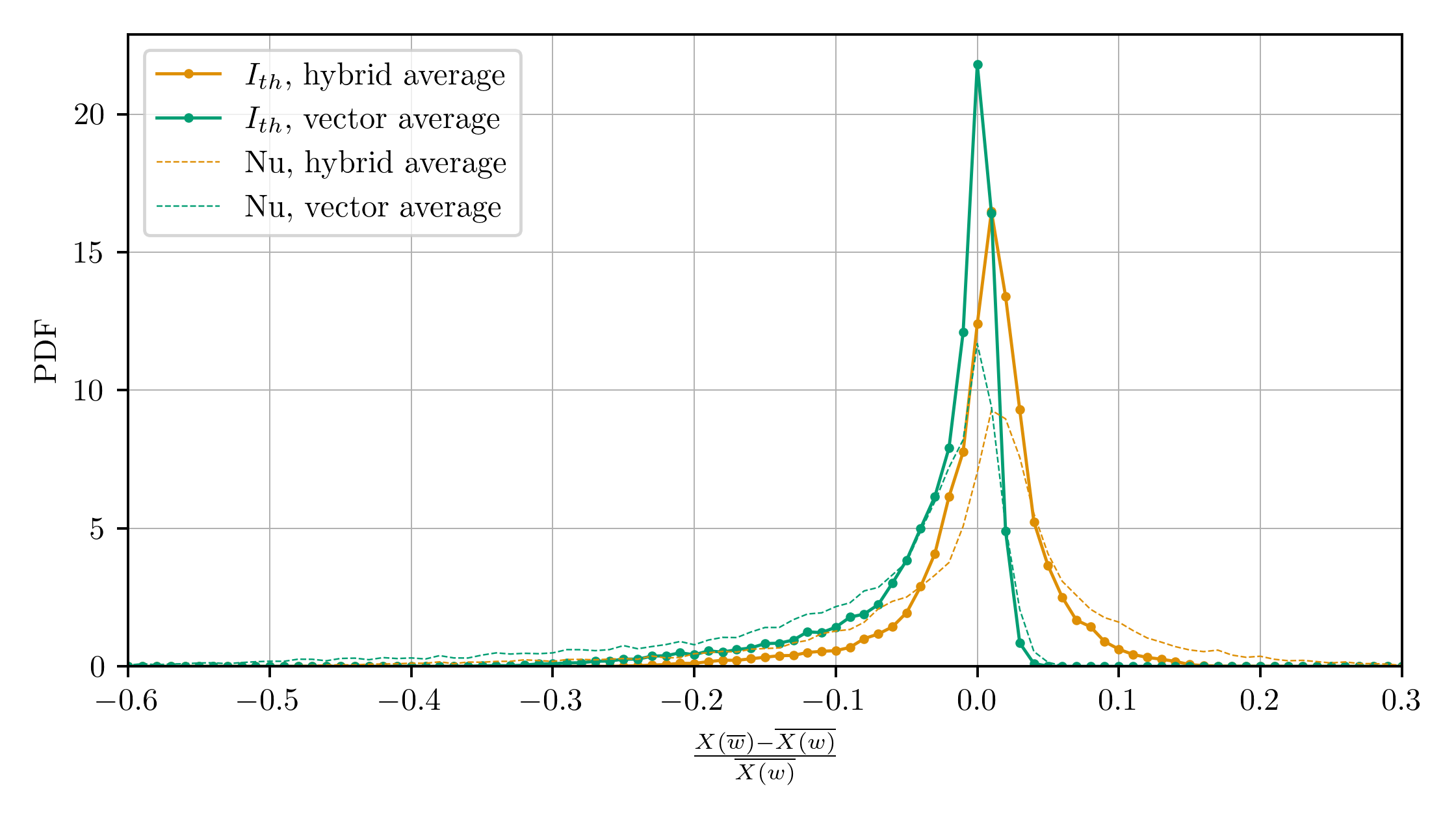}
	\end{center}
	\caption{Distribution of relative differences in Nusselt numbers and ampacities computed either from the averaged data $\nbr{Nu}(\overline{w})$ and $I_{th}(\overline{w})$ or the corresponding averaged Nusselt number, \Nuavg, and ampacity, \Iavg\ computed on the full resolution data. Averaging window size is \SI{5}{\minute}.}
	\label{fig:dtr_errorHist}
\end{figure}

Examples A and B were chosen from the best- and worst-case ends of the spectrum. Let us now examine the whole dataset to assess the overall effect.
Figure~\ref{fig:dtr_errorHist} shows the distribution of relative differences for Nusselt number and ampacity for both averaging methods and the whole dataset. 
With vector averaging, ampacity was higher than \Iavg around \SI{20}{\percent} of the time, but the highest difference was quite limited with the amount of time exceeding \SI{5}{\percent} difference being negligible. The majority of time, ampacity computed on averaged data was lower than \Iavg. The differences amounted to being higher than \SI{10}{\percent} in \SI{13}{\percent} of the time and higher than \SI{20}{\percent} in \SI{3}{\percent} of the time.

With hybrid averaging, the relative differences distribution is more symmetric, and underestimates the ampacity \SI{45}{\percent} of the time. For \SI{4}{\percent} of the time it underestimate \Iavg\ for more than \SI{10}{\percent} and for less than around \SI{0.5}{\percent} of the time it underestimates \Iavg\ for more than \SI{20}{\percent}. On the other hand, around \SI{2}{\percent} of the time ampacity overestimates \Iavg\ for more than \SI{10}{\percent}. 

The differences in Nusselt number show the same trend as differences in ampacity for both averaging methods, but with even more prominent distributions. In ampacity, the effect of averaging is damped by other heat terms. 

This leads us to the main takeaway of this paper: \textit{wind velocity averaging significantly impacts DTR calculations, and different averaging methods impact the results differently.} We will analyse this finding in more detail in the next sections.

\subsection{Angle dependency}\label{sec:angle-dependency}

\begin{figure}[H]
	\begin{center}
		\includegraphics[width=0.49\textwidth]{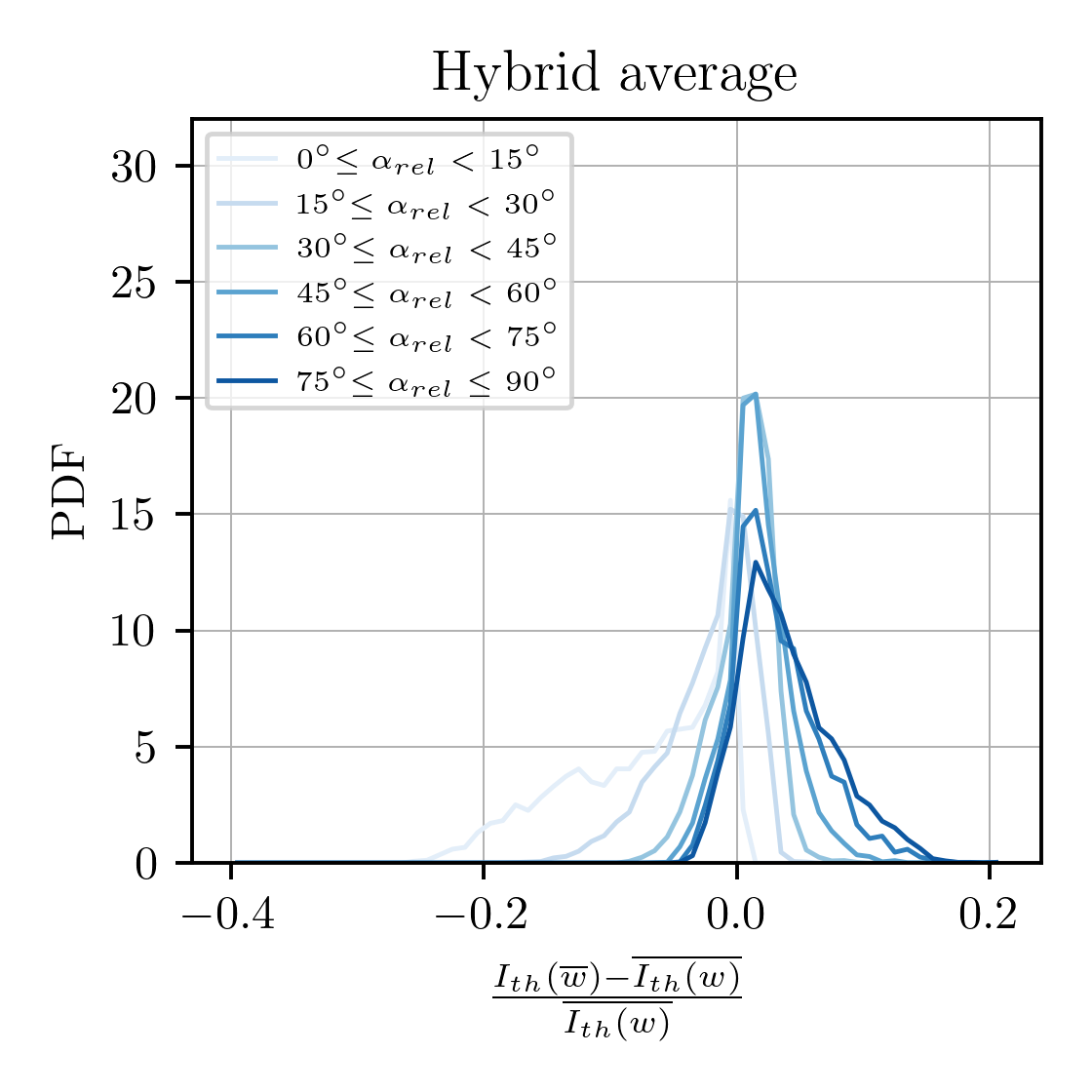}		
		\includegraphics[width=0.49\textwidth]{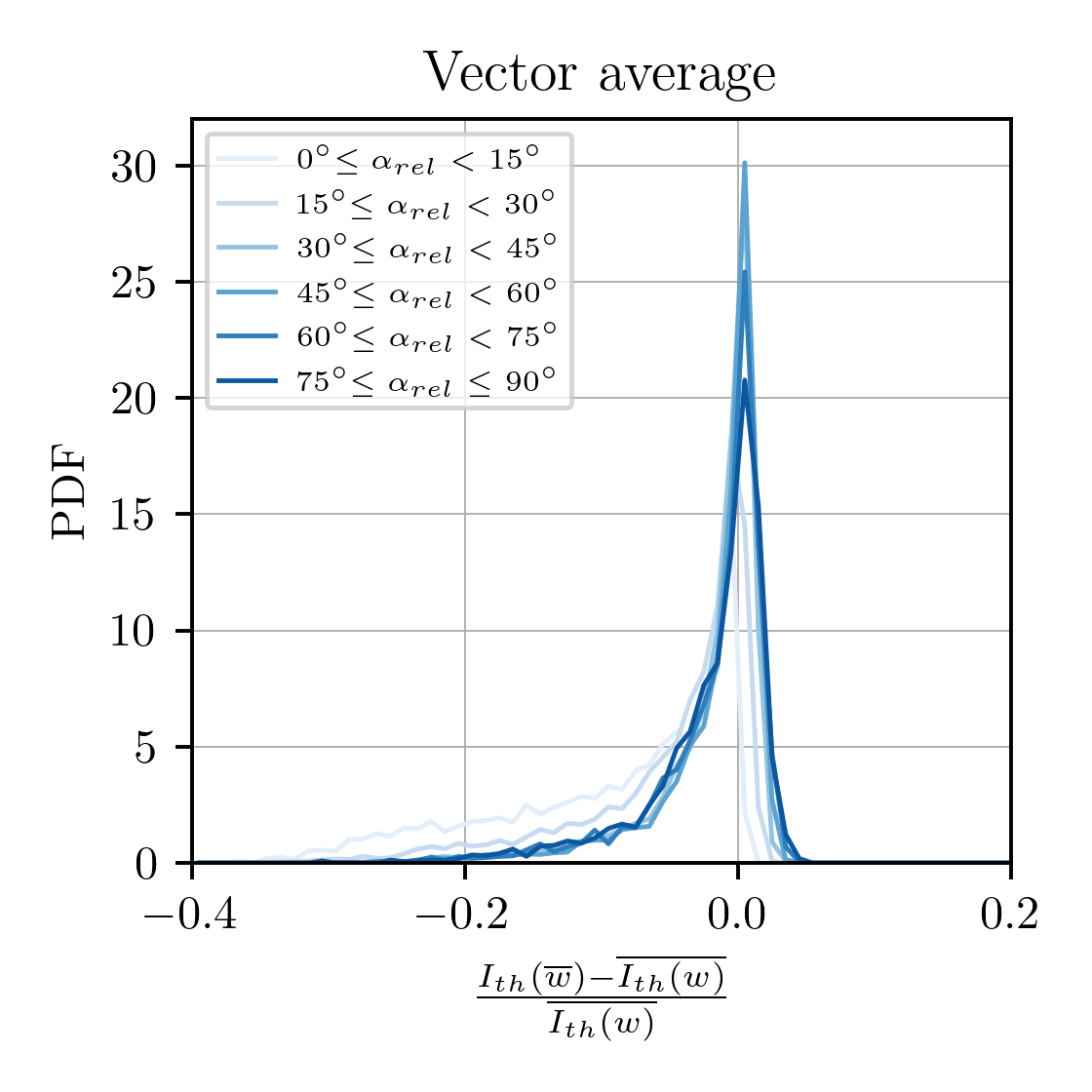}
	\end{center}
	\caption{Relative ampacity difference distributions for different relative angles for hybrid (left) and vector averaging (right).}
	\label{fig:analysisAngle_errorHist}
\end{figure}

We have seen from examples in \ref{sec:dtr} that the relative angle plays an important role, so let us take a look at how ampacity difference changes with the average relative angle. Figure~\ref{fig:analysisAngle_errorHist} shows the distribution of relative ampacity differences for \SI{5}{\minute} windows for both averaging methods, broken down by average relative angles. With vector averaging, the distributions appear heavily asymmetric, with \Iavg\ being underestimated the majority of time. The tails are the longest for small relative angles, while with the increasing relative angle, the amount of underestimated ampacities decreases and the distributions become narrower, with the peaks moving towards the right. At small relative angles under \ang{15}, ampacity based on averaged data underestimates the \Iavg\ for more than \SI{10}{\percent} over \SI{38}{\percent} of the time, and for more than \SI{20}{\percent} for around \SI{13}{\percent} of the time. At close-to-perpendicular wind where the highest percentage of time overestimates \Iavg, the amount of time that ampacity based on averaged data overestimates \Iavg\ for more than \SI{4}{\percent} are negligible. With hybrid averaging, the distributions become more symmetric. At lower relative angles, longer tails in the negative direction can be observed -- average-based ampacity underestimates \Iavg. With the increase in angle, the tails shift towards the right, with \Iavg\ being overestimated a significant proportion of time. Observing the numbers, the underestimated \Iavg\ has smaller absolute relative differences at close-to-parallel wind than with vector averaging. A bit over \SI{34}{\percent} of the time \Iavg\ is underestimated for more than \SI{10}{\percent}, and only around \SI{4}{\percent} of the time it is by \SI{20}{\percent} or more. On the other hand, around \SI{10}{\percent} of the time \Iavg\ is overestimated for more than \SI{10}{\percent} at the close-to-perpendicular wind. 

Intuitively, the observed shift from underestimating \Iavg\ to overestimating it with the increase in the angle is expected. If we consider a case with parallel wind, where the average angle is \ang{0}, any variability in wind direction increases the cooling. Therefore with averaging, we are looking at the worst-case scenario, so we expect that \Iavg\ will be higher than that. On the other hand, with average wind at \ang{90}, the averaging results in the best scenario, and any variability in wind direction decreases the cooling. For the relative angles in between, the variability in wind direction means there are times with better and times with worse cooling, with the effect partially cancelling itself out, making the tails in the distribution of differences less pronounced. Note that the transition between the two extremes is not linear. This can be seen from as Eq. \ref{eq:Cigre_NuAngle}, which includes sine dependence, as well as a step-change in parameters.

\section{Limit cases}\label{sec:limit-cases}

Establishing the relative angle as an important factor, let us analyse two limit cases where the relative angle is \ang{0} and \ang{90}. On one hand, the case with parallel wind is of special interest to TSOs, as in practice, the limiting spans are often the spans where the wind is parallel to the line. Therefore, the accuracy of calculated ampacities could have a significant effect on line operation in this regime. On the other hand, the case with perpendicular wind is the best-case scenario for convective cooling, and as discussed earlier, this is the regime where it is most likely that the ampacity on average data overestimates \Iavg. Therefore, this regime requires additional analysis to help us understand the risk that averaging brings. As the span of the available data for analysis is relatively short (\SI{3}{months}), and skewed with the uneven wind direction distribution, we must augment the data to support further analysis.

We start from the static property of the ampacity. Since ampacity is defined for static conditions, it can be computed for each time window in isolation and neighbouring time windows do not impact it. To control the wind angle, we can rotate it uniformly without affecting its variability. Therefore, we will split the data into \SI{5}{\minute} windows and independently rotate the wind within each window to analyse the effects of relative wind angle on DTR.

\subsection{Parallel wind}

\begin{figure}[H]
	\begin{center}
		\includegraphics[width=\textwidth]{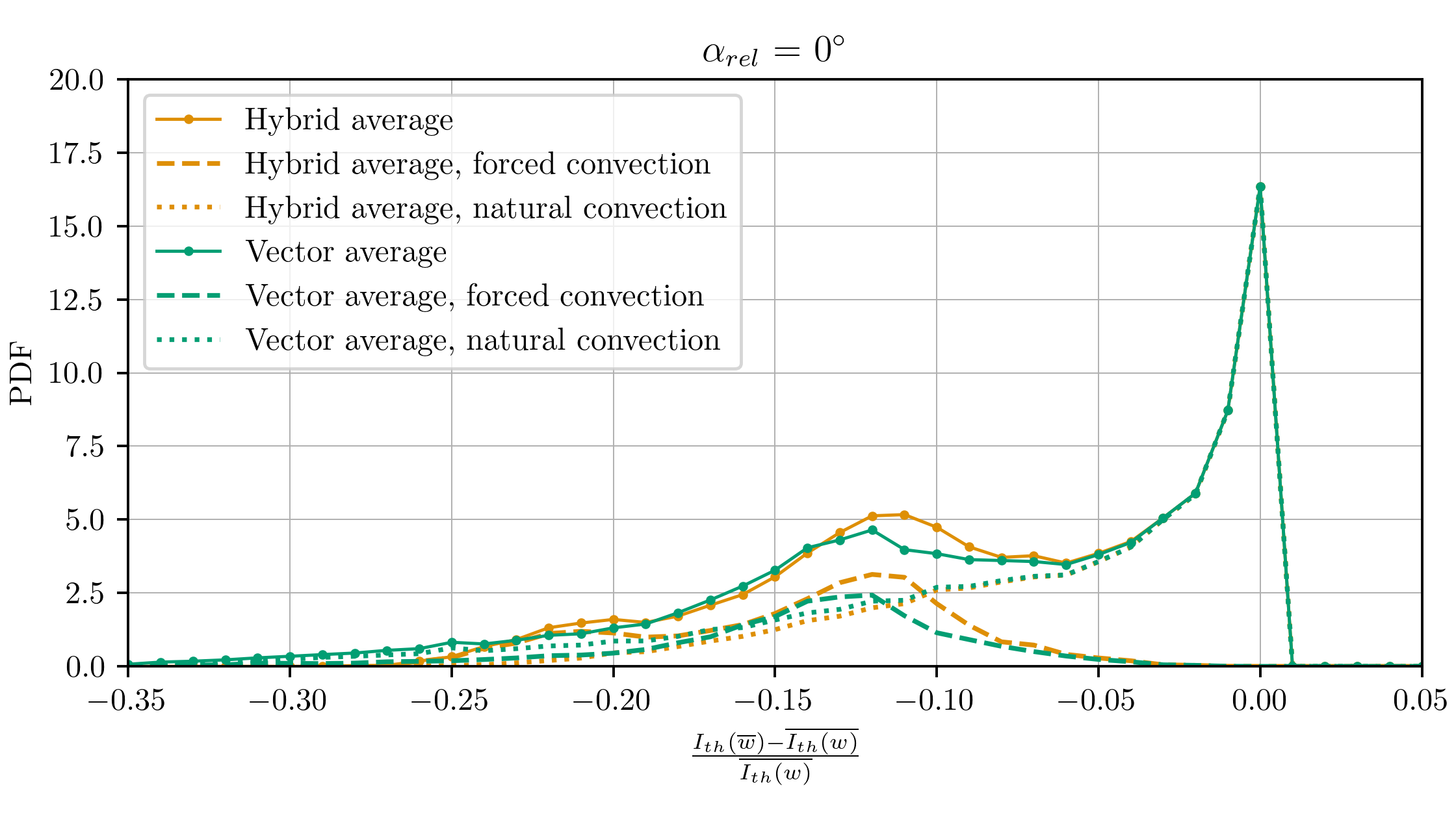}	
	\end{center}
	\caption{Relative ampacity differences for $\alpha_{rel}=\ang{0}$ for hybrid and vector averaging, with the breakdown in forced and natural convection regimes.}
	\label{fig:limitCase0_errorHist}
\end{figure}

In the first case, we align all the average wind angles with the line. Figure~\ref{fig:limitCase0_errorHist} shows the distribution of relative ampacity differences for both averaging methods. As mentioned above, convective cooling is the least effective at $\alpha_{rel}=\ang{0}$, so any variability in wind direction increases the cooling. This is why ampacities on averaged data underestimate \Iavg. As mentioned in section \ref{sec:dtr}, CIGRE differentiates between forced and natural convection regimes, therefore in Figure~\ref{fig:limitCase0_errorHist} segmentation between the regimes is also shown.

Note that for any window classified as one regime based on \SI{5}{\minute} window, some \SI{1}{\second} data within that window may actually fall in the other regime. We see that the spike at around \SI{0}{} is due to natural convection, while the forced convection distribution has a peak between \SI{-0.2}{} and \SI{-0.1}{}. The overall shapes of PDFs are similar for both averaging methods, with the distribution for vector averaging having longer tails.

\begin{figure}[H]
	\begin{center}
		\includegraphics[width=0.49\textwidth]{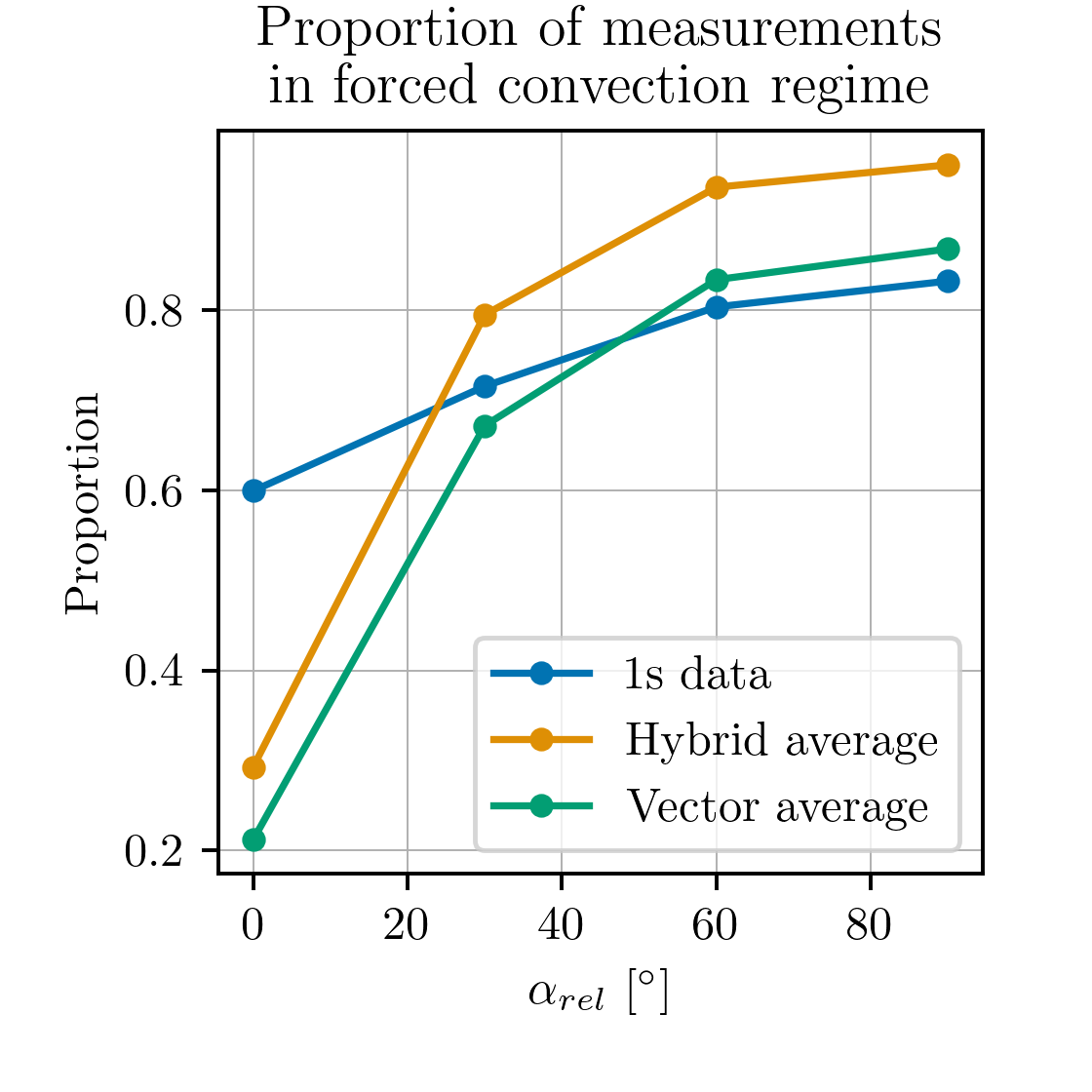}
		\includegraphics[width=0.49\textwidth]{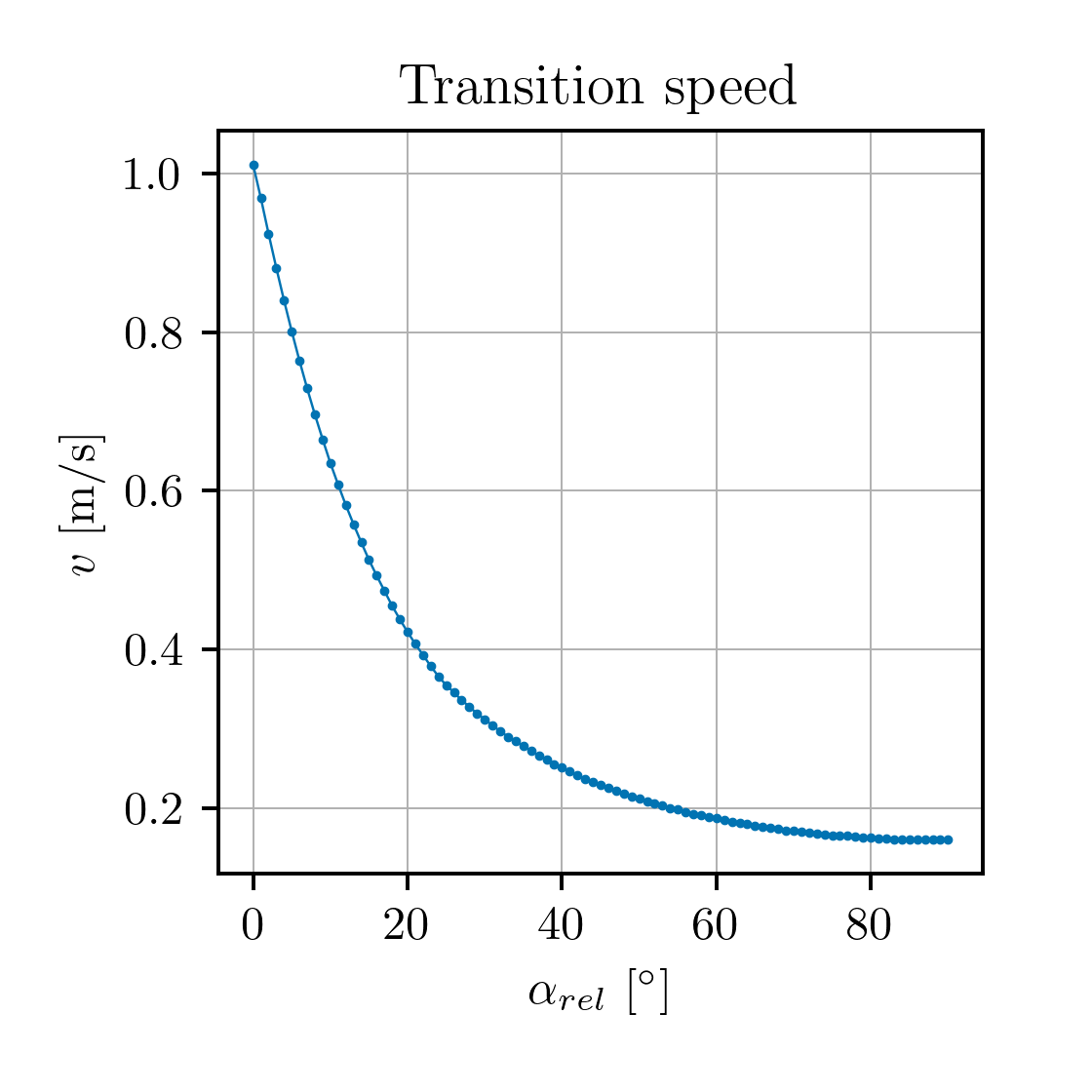}		
	\end{center}
	\caption{Left: Proportion of cases that fall into forced convection regime depending on the relative direction of the wind for \SI{1}{\second} data and both averaging methods. Right: The speed at which transition between forced and natural convection occurs for $T_s=\SI{80}{\degreeCelsius}$ and $T_a=\SI{15}{\degreeCelsius}$, depending on the relative wind angle.}
	\label{fig:limitCase_forcedConvection}
\end{figure}

Another thing we note is that a substantial proportion of samples lies within the natural convection regime. Figure~\ref{fig:limitCase_forcedConvection} (left) shows the proportion of samples with dominant forced convection regime for \SI{1}{\second} data and \SI{5}{\minute} averaged data, as a function of the relative angle. We notice that the proportion rises with the average angle, and that at \ang{0}, the proportion of samples with dominant forced convection is much lower for averaged data than for \SI{1}{\second} data.

This can be explained with Figure~\ref{fig:limitCase_forcedConvection}, which shows the speed at which the transition between natural and forced convection occurs (we will refer to it as transition speed) at the selected conditions, i.e. the selected constant air temperature and maximum conductor temperature. The transition speed decreases with the angle, and has the steepest slope around \ang{0}, so any variability in wind direction can cause the \SI{1}{\second} data sample to fall into a different regime than the averaged data sample. It can also be clearly seen from Figure~\ref{fig:limitCase_forcedConvection} (right) that for all angles, more data falls into the natural convection regime with vector averaging, compared to hybrid averaging. This is because for windows with large variability in the direction, the vector averaging dampens the speed.

\begin{figure}[H]
	\begin{center}
		\includegraphics[width=0.49\textwidth]{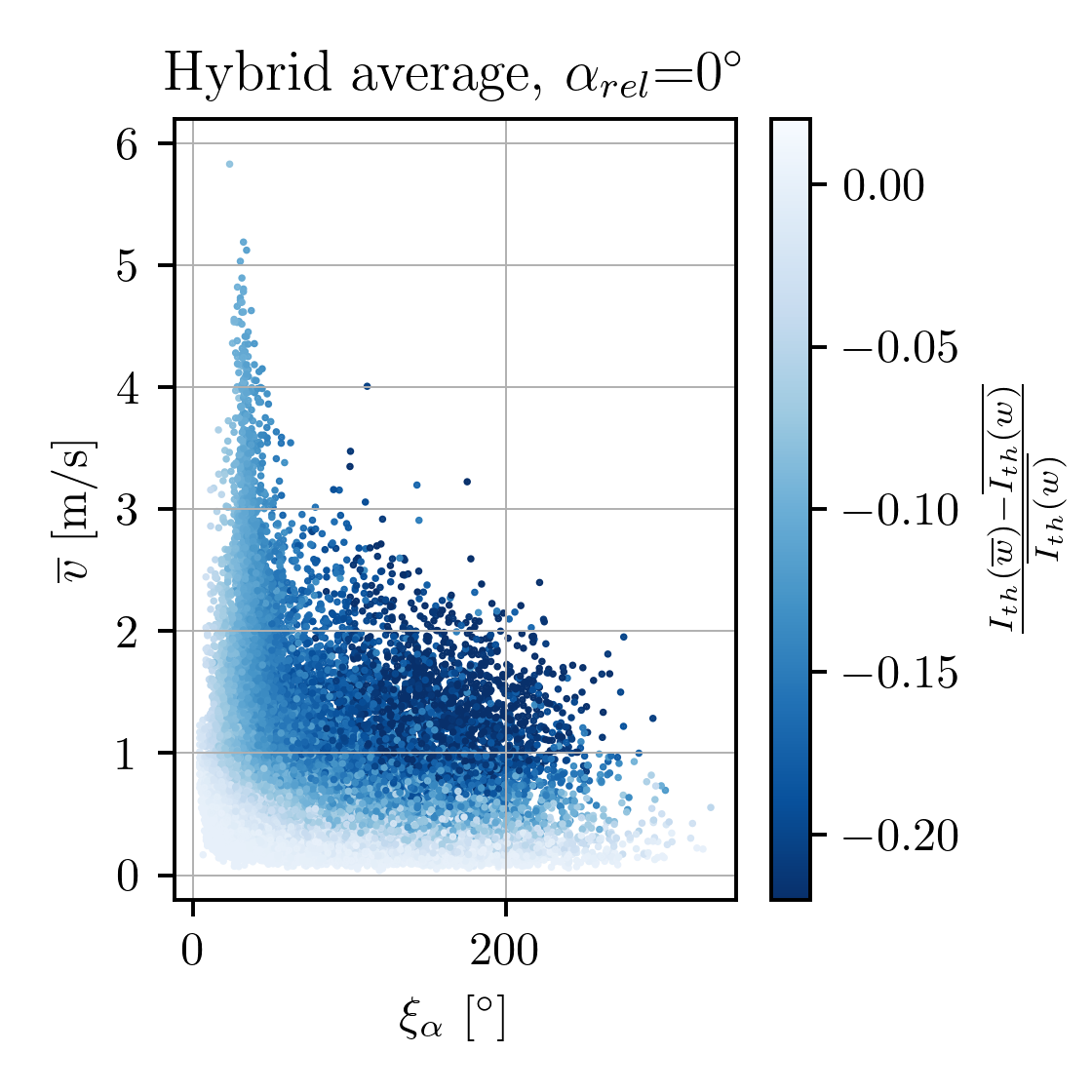}	
		\includegraphics[width=0.49\textwidth]{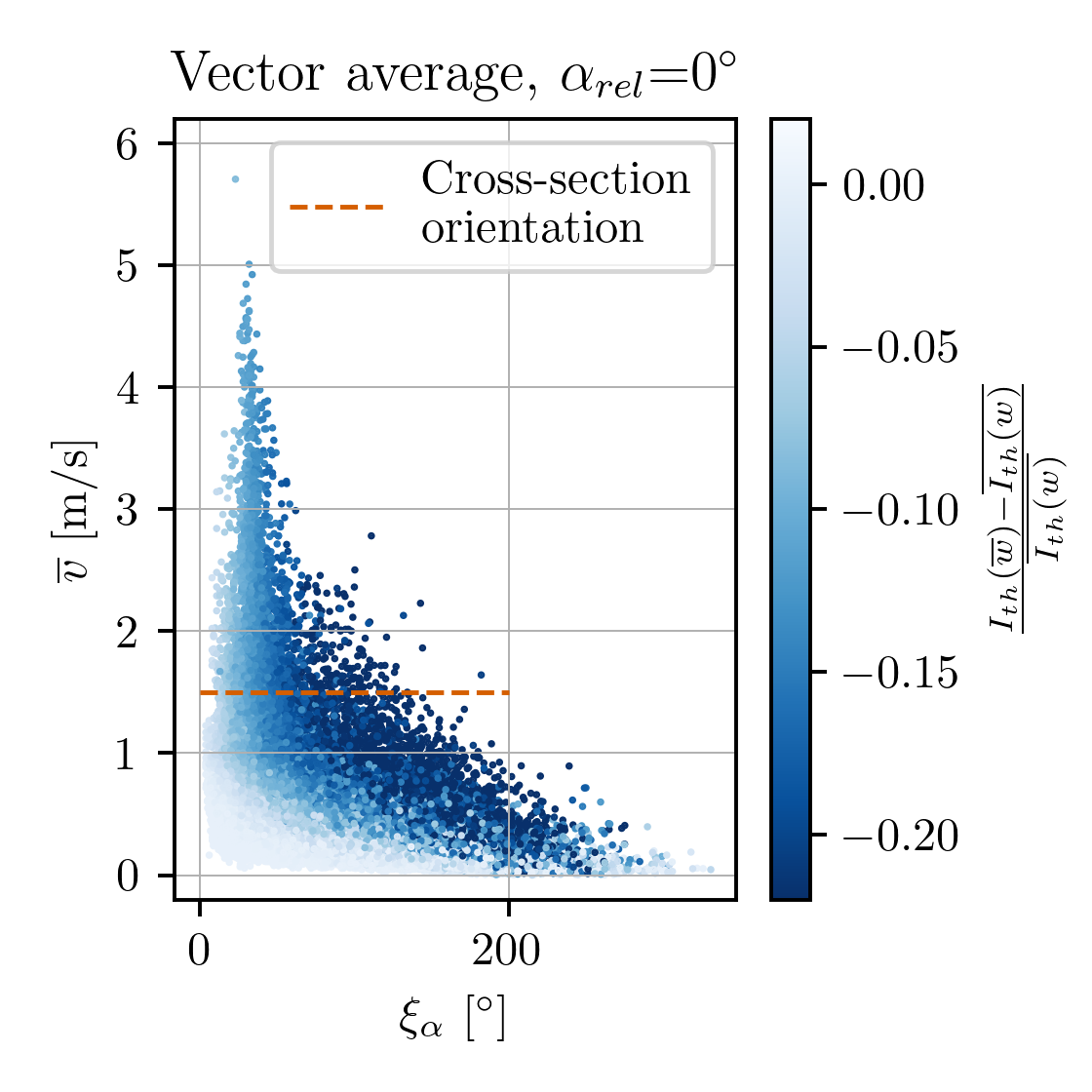}	
	\end{center}
	\caption{Relative ampacity difference in parallel wind case as a function of $\xi_{\alpha}$ and $v$ for hybrid (left) and vector averaging (right). Note that the scale is truncated for clarity.}
	\label{fig:limitCase0_errorScatter}
\end{figure}

Figure~\ref{fig:limitCase0_errorHist} established that there is a difference between ampacity computed on averaged data versus computed as \Iavg . Let us take a look at how this difference depends on wind variability. Figure~\ref{fig:limitCase0_errorScatter} shows the relation between the aforementioned difference, the wind direction variability, and the mean wind speed. The latter was selected in favour of wind speed variability since it represents a readily available measurement and is at the same time linearly correlated to wind speed variability (Figure~\ref{fig:wind_averagingSpeeds}). Thus, if the figure featured $\xi_{\alpha}$ in place of $\overline{v}$, the plots would remain the same in overall form and relationship. 

\begin{figure}[H]
	\begin{center}
		\includegraphics[width=0.49\textwidth]{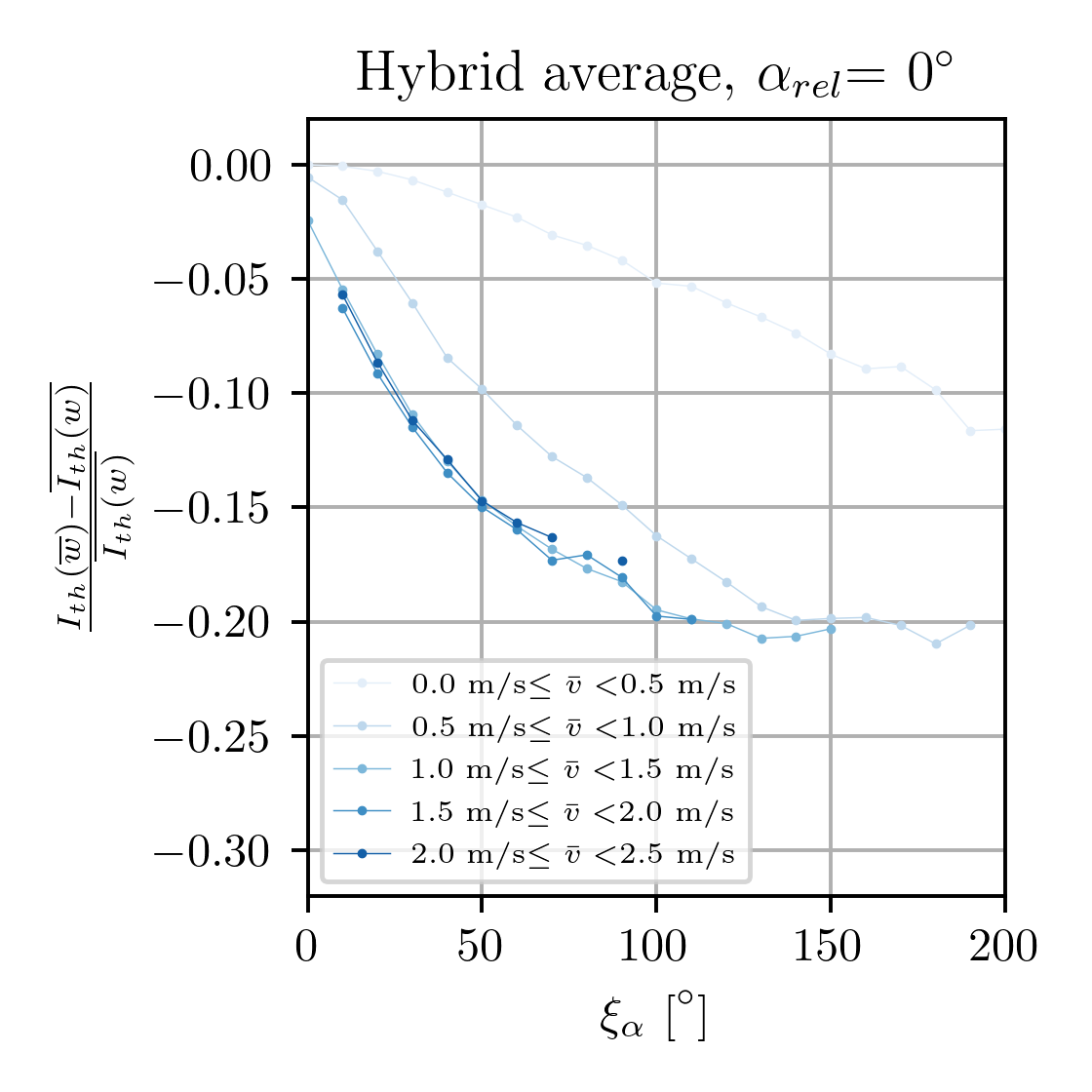}	
		\includegraphics[width=0.49\textwidth]{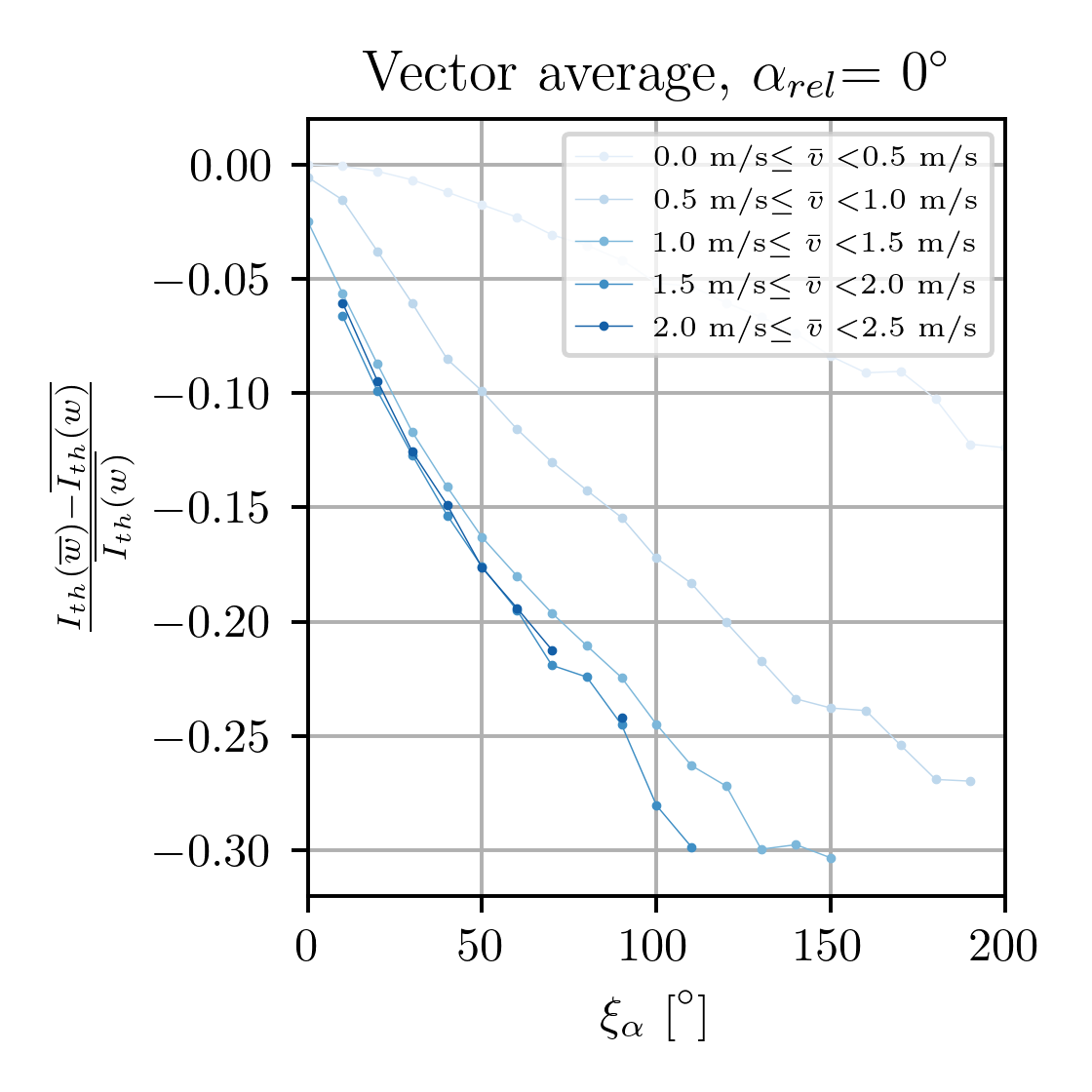}	
	\end{center}
	\caption{The cross-section plots showing relative ampacity difference in parallel wind case as a function of $\xi_{\alpha}$ for hybrid (left) and vector averaging (right) for several wind speed intervals.}
	\label{fig:limitCase0_errorCross}
\end{figure}

Figure~\ref{fig:limitCase0_errorScatter} shows how the difference between ampacity computed on averaged data versus computed as \Iavg\ depends on the wind speed and wind direction variability. Note that the colour scale is truncated for clarity. We can see that both plots have similar shapes. For both types of averaging, the absolute minimum relative ampacity differences are obtained at low wind speeds and low variabilities in wind direction. The relative difference becomes larger (in the negative direction) with the increase of both parameters. Figure~\ref{fig:limitCase0_errorCross} shows the cross-sections of the scatter plot for different speed intervals. We can see that for all speed ranges, with increasing $\xi_{\alpha}$, the ampacity difference first gets more negative, and then reaches a plateau. A plateau is expected, as ampacity is dependent on the relative angle, so there is a point when the increase in angle variability does not introduce any new relative angles. The location of the plateau depends on the parameter $p$ in our definition of variability. We saw in Figure~\ref{fig:limitCase_forcedConvection} that for approximately $v>\SI{1}{\ms}$, forced convection outweighs the natural convection, which is reflected in Figure~\ref{fig:limitCase0_errorCross} through the minimal differences between the lines that correspond with the forced convection regime.

The observed dependency could serve as a measure for evaluation of the discrepancy in the ampacity of the averaged data compared to the \Iavg, provided that the information about wind variability inside the window is available. This could potentially be of great interest in practice, as it effectively means that the majority of time when operating in parallel wind, the ampacity is underestimated. In the observed data, if wind variability was accounted for properly, the ampacity estimate could be raised considerably. However, further analysis and a better model are needed before it can be adopted by transmission operators.

\subsection{Perpendicular wind}

\begin{figure}[H]
	\begin{center}
		\includegraphics[width=\textwidth]{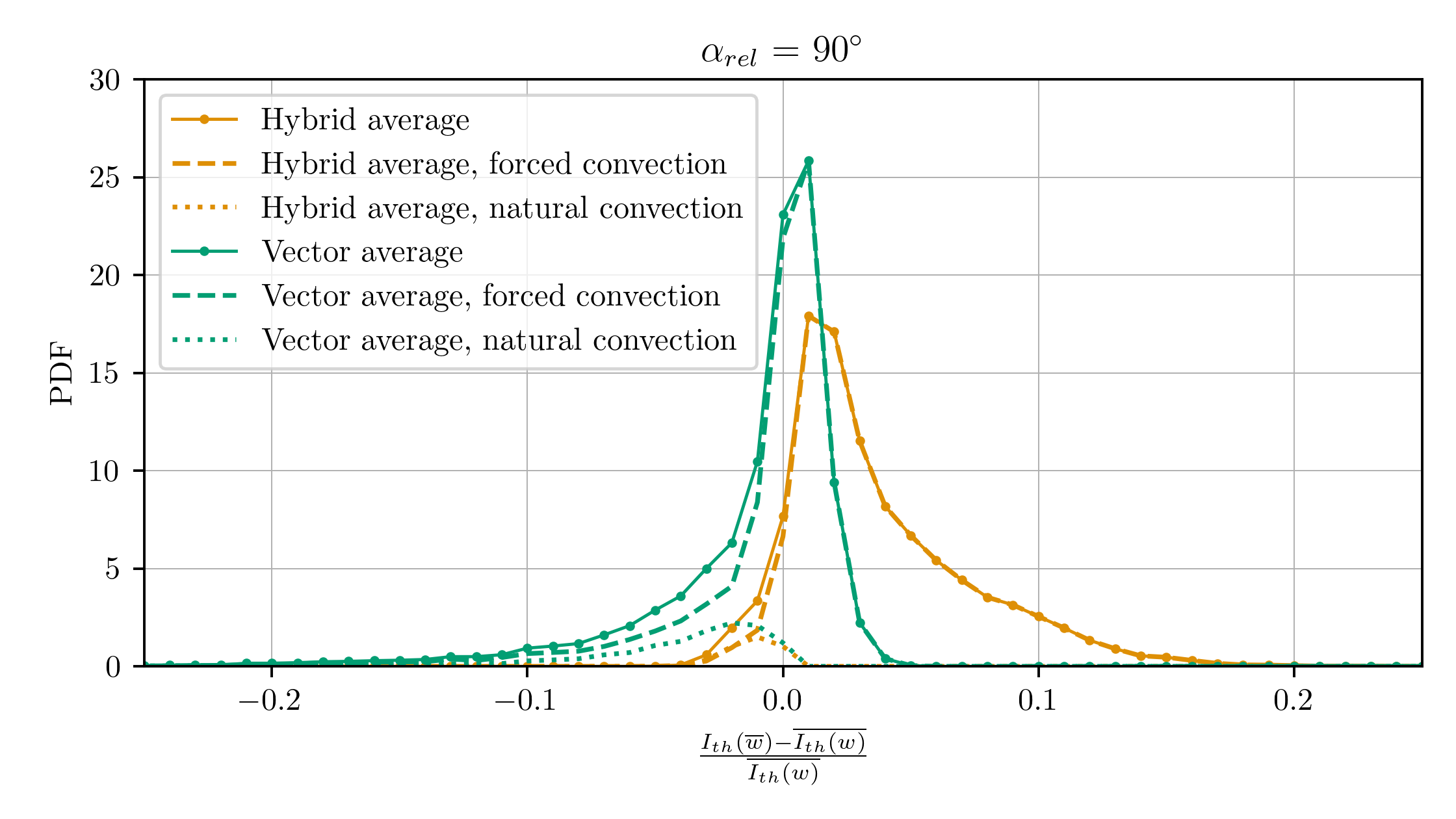}	
	\end{center}
	\caption{Relative ampacity differences for $\alpha_{rel}=\ang{90}$ for hybrid and vector averaging, with the breakdown in forced and natural convection regimes.}
	\label{fig:limitCase90_errorHist}
\end{figure}

Finally, let us take a look at the case with perpendicular wind. Figure~\ref{fig:limitCase90_errorHist} shows the relative ampacity difference distribution for both averaging methods. Convective cooling at \ang{90} is the most effective, so any variability in wind angle lowers the convective cooling. This is why averaging overpredicts the ampacity for a considerable amount of time, which is especially true for hybrid averaging, where the scalar average of the speed does not dampen the amplitude with increased variability. On the other hand, vector averaging does dampen the amplitude, which is why in this case, ampacity is underpredicted for a considerable amount of time. Note that the proportion of the natural convection regime is much lower than with parallel wind, which is in accordance with the lower transition speed from Figure~\ref{fig:limitCase_forcedConvection} (right).

\begin{figure}[H]
	\begin{center}
		\includegraphics[width=0.49\textwidth]{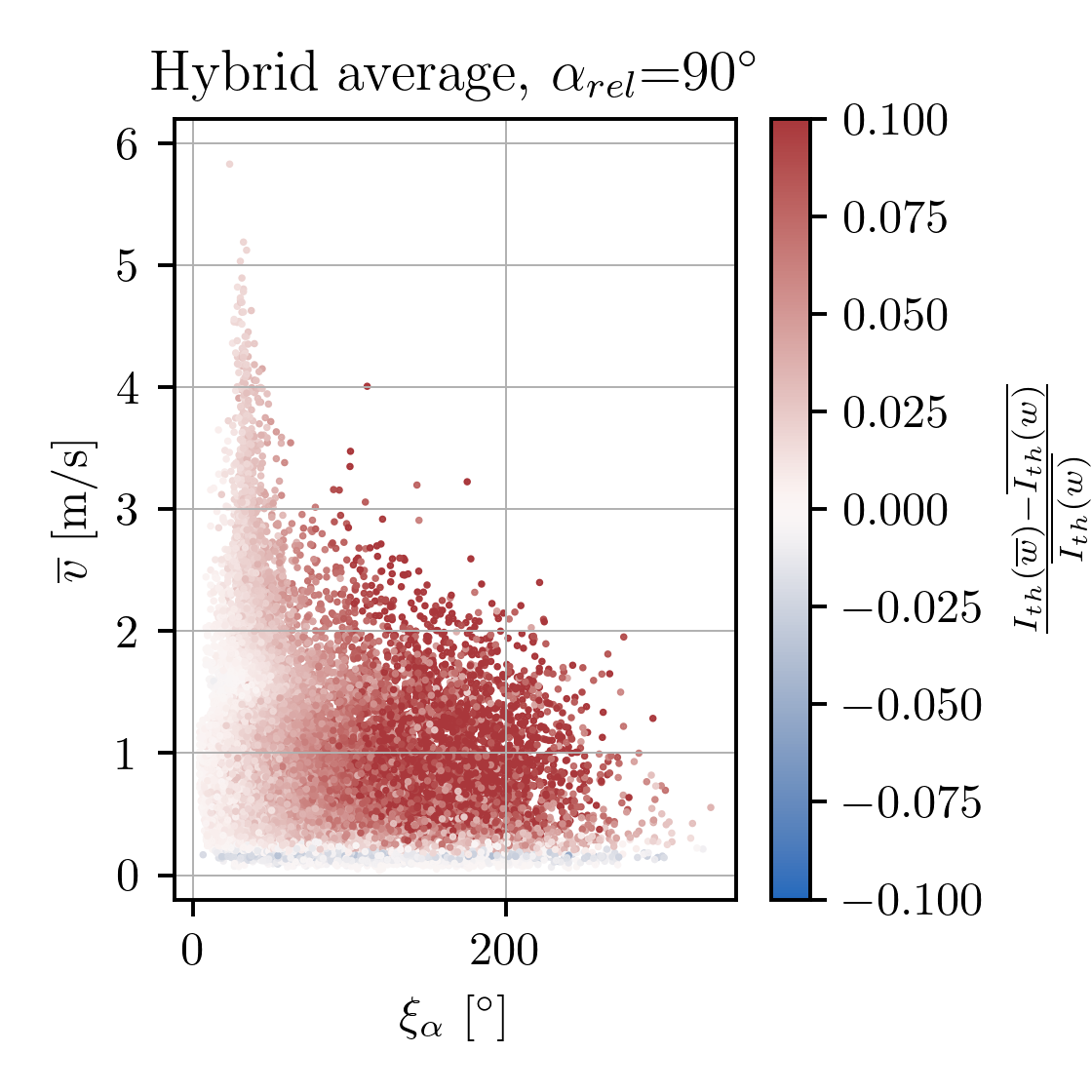}	
		\includegraphics[width=0.49\textwidth]{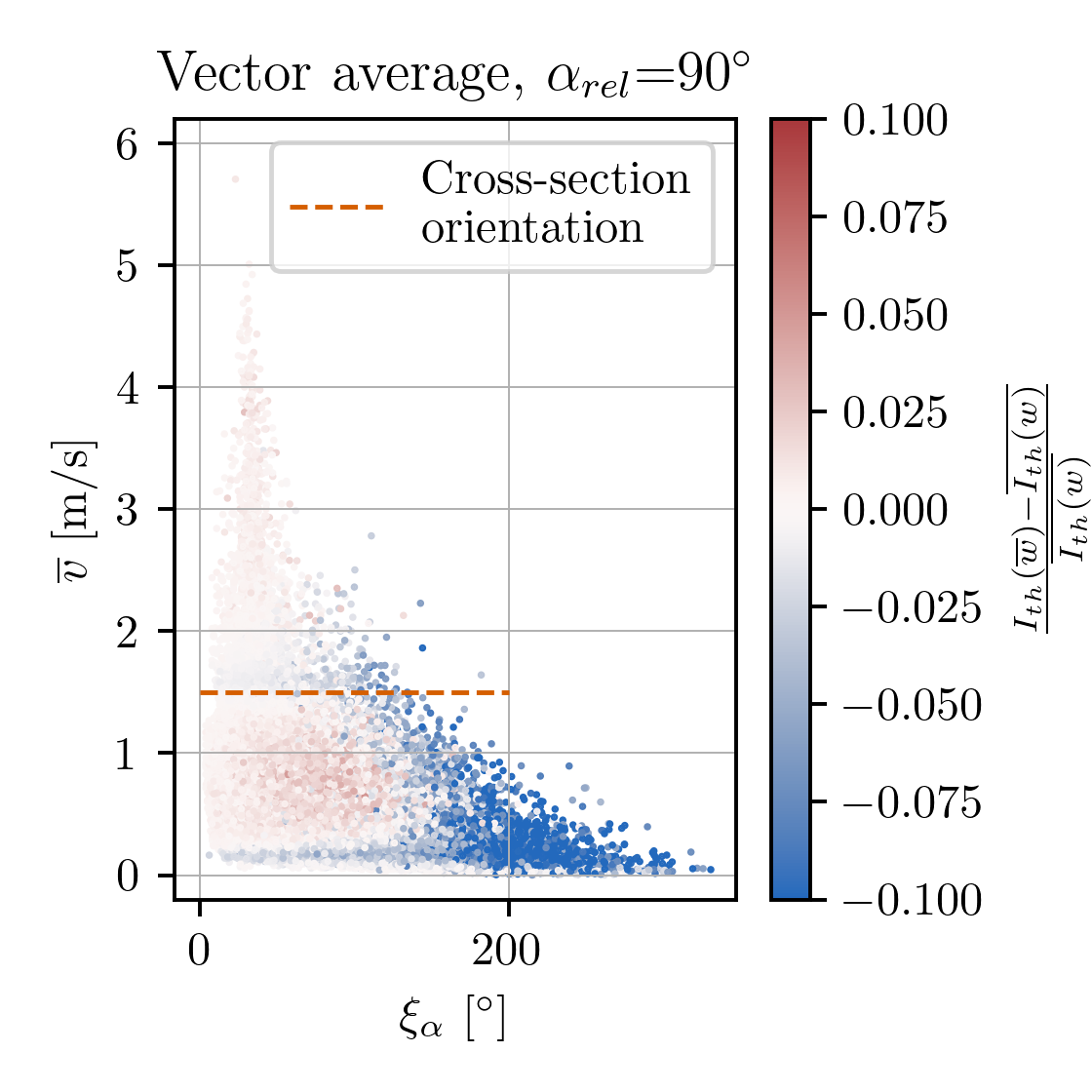}	
	\end{center}
	\caption{Relative ampacity difference in perpendicular wind case as a function of $\xi_{\alpha}$ and $v$ for hybrid (left) and vector averaging (right). Note that the scale is truncated for clarity.}
	\label{fig:limitCase90_errorScatter}
\end{figure}

\begin{figure}[H]
	\begin{center}
		\includegraphics[width=0.49\textwidth]{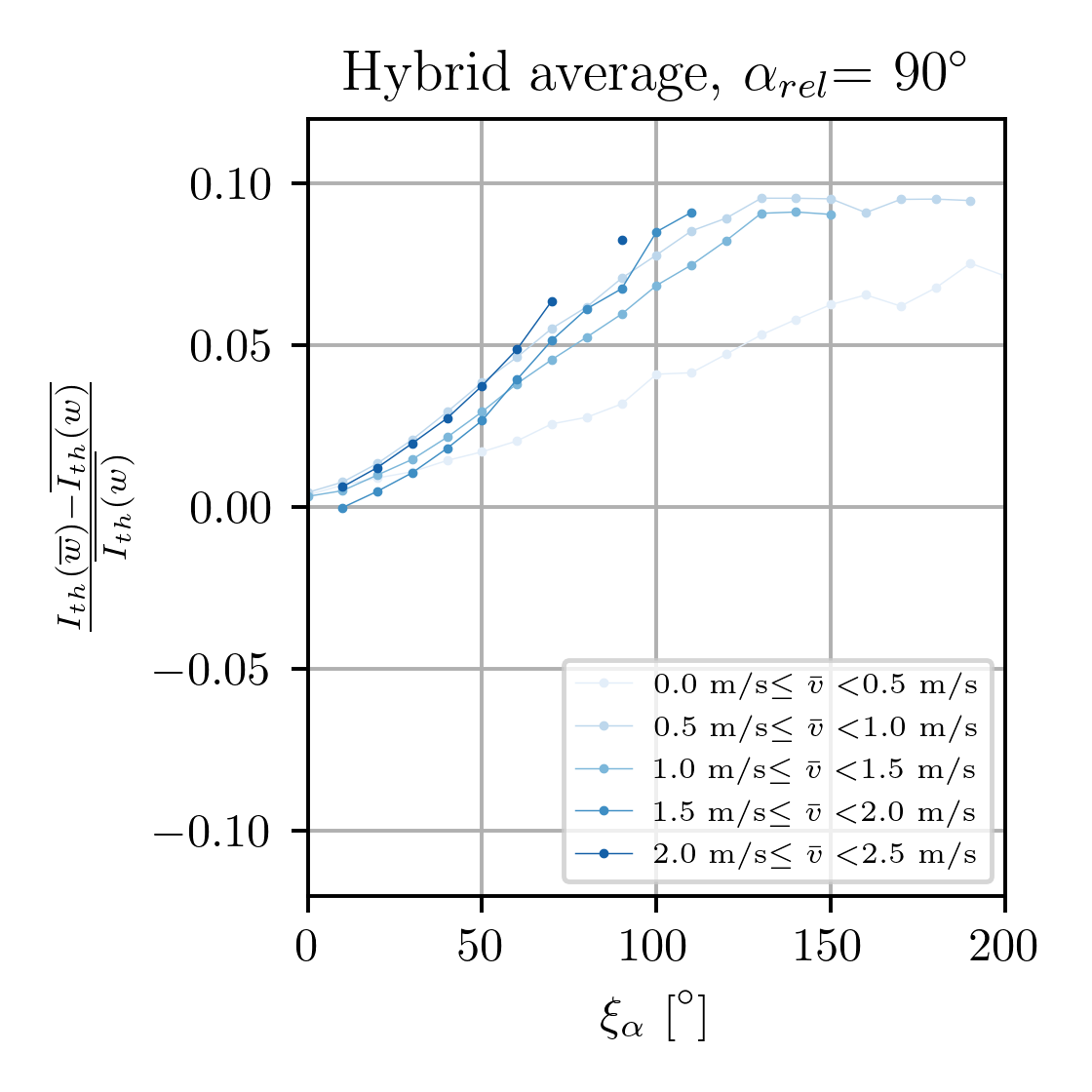}	
		\includegraphics[width=0.49\textwidth]{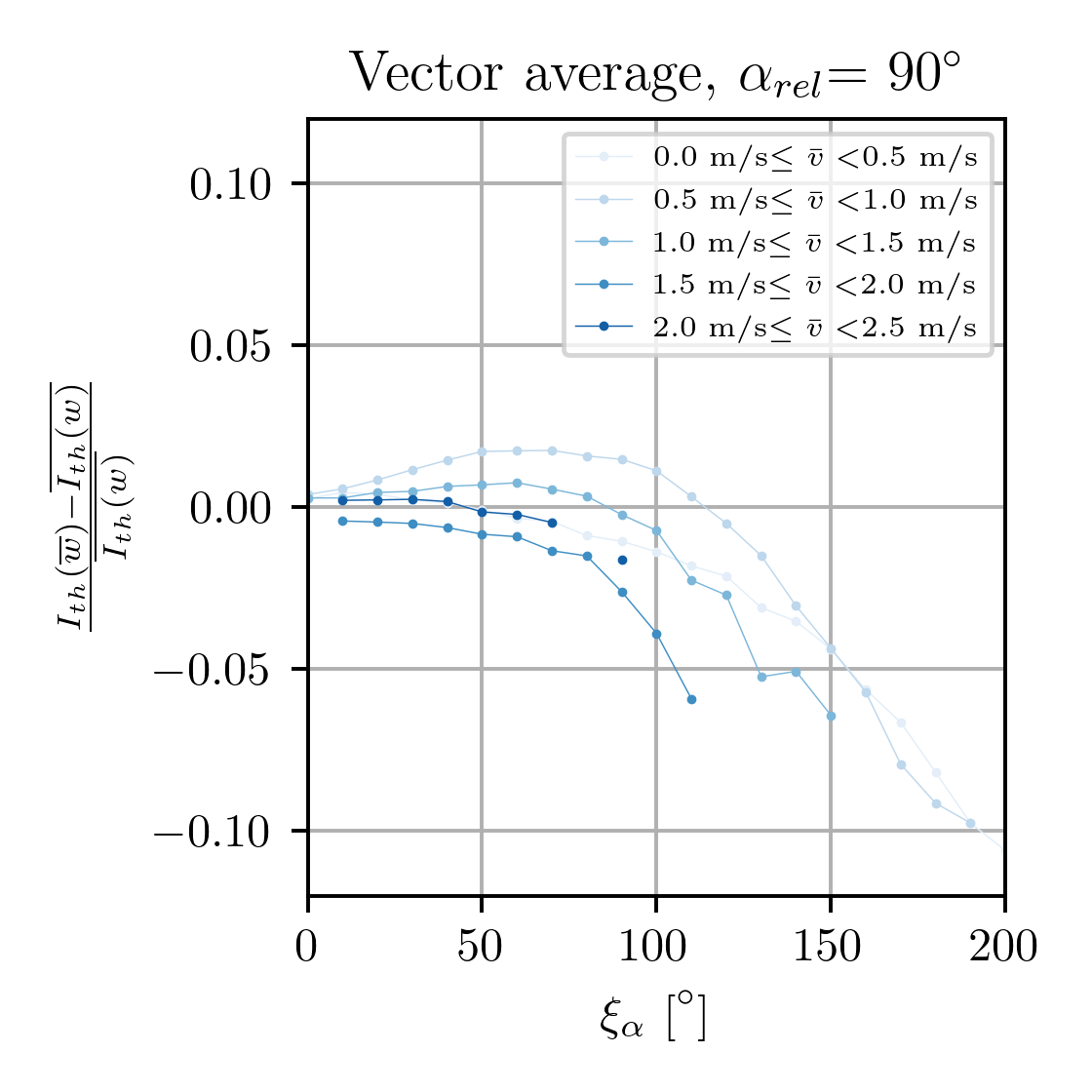}	
	\end{center}
	\caption{The cross-section plots showing relative ampacity difference in perpendicular wind case as a function of $\xi_{\alpha}$ for hybrid (left) and vector averaging (right) for several wind speed intervals.}
	\label{fig:limitCase90_errorCross}
\end{figure}

Next, Figure~\ref{fig:limitCase90_errorScatter} shows how relative ampacity difference depends on wind speed and $\xi_{\alpha}$. For the hybrid average, the shape of the graph is reversed in comparison with the parallel wind case. The minimum differences are again obtained at low values of both parameters and then increase with the increase in either parameter. The sign of the relative difference is different than in the parallel wind case, as is expected, and the maximum amplitude is smaller, which is also expected, because of the sine dependency on the relative angle in Nusselt number calculations in \ref{eq:Cigre_NuAngle}. We can see this effect also in Figure~\ref{fig:limitCase90_errorCross} (left), where we get the nearly linear increase of the relative difference followed by a plateau, mirroring the case with parallel wind.

For vector averaging, the scatter plots are qualitatively different. In parallel wind case, we have two effects that caused the ampacity on averaged data to be lower than \Iavg, namely vector averaging which dampens the wind amplitude, and the ineffectiveness of cooling at $\alpha_{rel}=\ang{0}$. In perpendicular wind case, the two effects work in opposite directions: the dampening of amplitude still lowers ampacity, but at $\alpha_{rel}=\ang{90}$, cooling by convection is the most effective, thus raising ampacity. The different functional shapes of the two methods can also be seen if we look at the cross-sections in Figure~\ref{fig:limitCase90_errorCross}. For vector averaging (right), a plateau appears at low $\xi_{\alpha}$, where the two effects cancel each other out, and starting from around $\xi_{\alpha}=\ang{100}$, the vector averaged speed dampening becomes dominant, resulting in the similar (near-linear) decrease as in the parallel wind case.

In light of these qualitatively different results, the choice of the averaging method becomes even more important in perpendicular wind case. While hybrid averaging seemed like a better choice in the previous sections, that is not the case here, as overpredicting ampacity could lead to unsafe operation.

\section{Conclusions}

In this paper, we analysed the \SI{1}{\second} temporal resolution wind measurements for a power line location in Slovenia. We looked at \SI{5}{\minute} averaging window and two averaging methods, vector averaging and hybrid averaging, and analysed the variability of wind speed and direction within the windows. We noticed the variability in wind speed increased with the speed, which is in agreement with the literature. Variability in wind direction takes up the whole spectrum for low wind speed, and with the increase of wind speed, it approaches a plateau of \ang{35} -- this is partly in agreement with the notion that stronger winds have better-defined direction (and lower variability) on one hand, while the variability increases with due to turbulence~\cite{cigre, raichle2009wind}. Further investigation of different locations is needed to determine whether this observation is universal, or location-specific. On the other hand, the dependence of the variability of wind speed and direction on the wind direction is definitely location-specific.

We performed DTR simulations on the available data and presented cases where both averaging methods give results similar to the data with higher temporal resolution, and cases where the averaging affects the results significantly. We compared the results of the high-resolution data and averaged data. Looking at the 3-month period, averaging has a measurable effect on the DTR results, and the effect depends on the relative wind angle. We examined two limiting cases, with the wind parallel and perpendicular to the line. In the case of parallel wind, using measurements with higher temporal resolution significantly increases the ampacity for both averaging methods. Vector averaging gives lower ampacities than scalar averaging, so TSOs using vector averaging can benefit more from the use of higher temporal resolutions. In the case of perpendicular wind, averaged data may sometimes overpredict the ampacity, so averaging should be done with caution. This effect is much more pronounced in hybrid averaging than in vector averaging.

Furthermore, the ampacity difference can be estimated if the variability in wind direction and the average wind speed are known. The ampacity differences found in this study are valid for the characteristic wind on the observed location, and the chosen limit temperature and weather parameters. Further studies with more site diversity and different weather conditions are needed before the findings can be generalised.

The main shortcoming of this study is applying established DTR models to short time scales, where we might be pushing the envelope of their validity. Further investigation is needed into the relations between Reynolds number and Nusselt number, especially with varying wind and turbulence. In this study, we characterised wind variability as a scalar number. Further analysis can be done of wind speed and direction distributions, including analysis of the distribution shapes, which might render additional insights. A DTR model to account for wind variability could be investigated in the future.

The main takeaway of this paper is that averaging wind data significantly affects the DTR results, and that the choice of the averaging method matters. Especially in the case of parallel wind, higher resolution data gives higher ampacity, which could be of interest to TSOs in case of limiting spans.

\section*{Appendix}\label{Appendix}
\renewcommand{\thesubsection}{\Alph{subsection}}
\subsection{Averaging window length}
\begin{figure}[H]
	\begin{center}
		\includegraphics[width=0.49\textwidth]{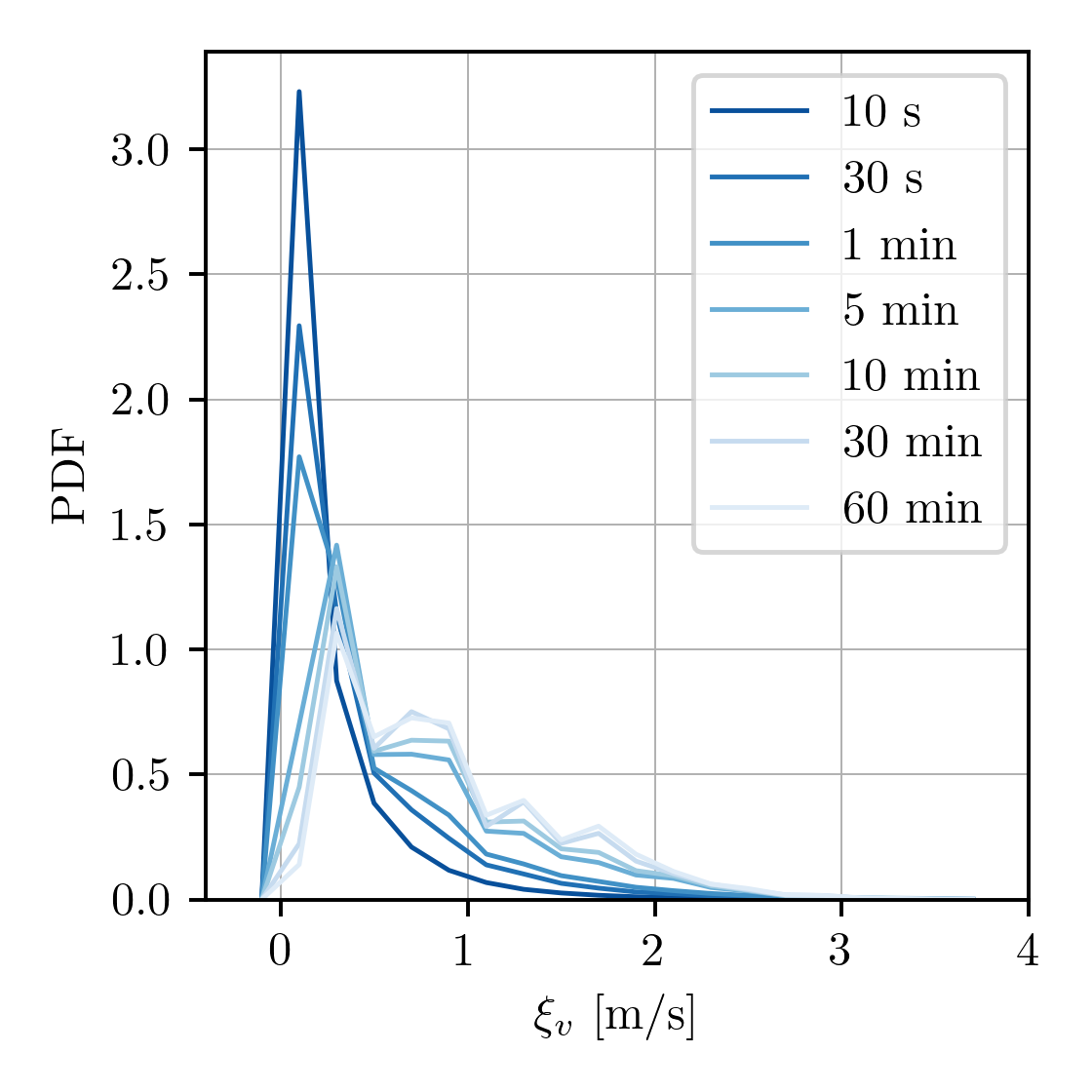}		
		\includegraphics[width=0.49\textwidth]{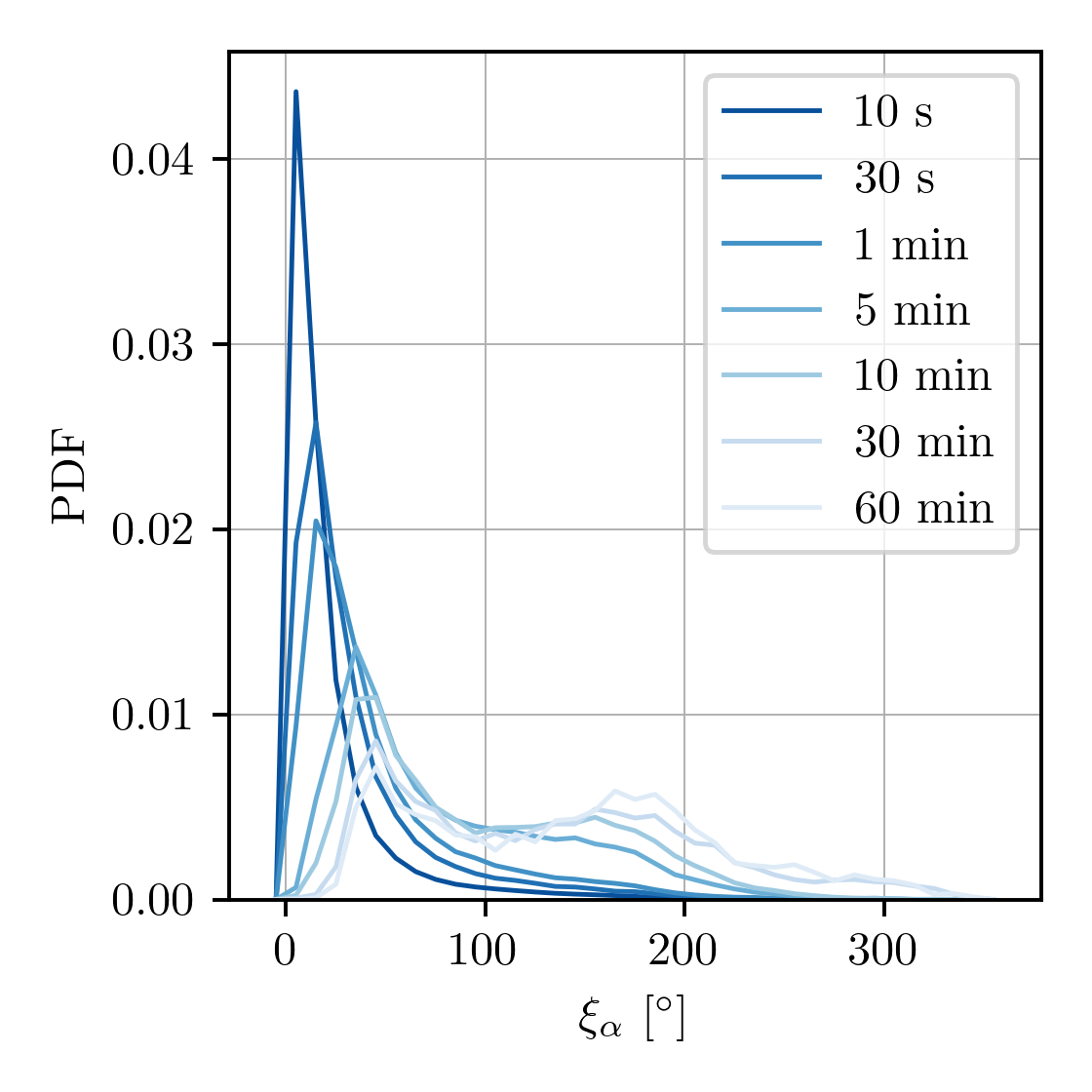}
	\end{center}
	\caption{Distribution of $\xi_v$ (left) and $\xi_{\alpha}$ (right) for all of the observed data, for different lengths of the averaging window.}
	\label{fig:analysisWindow_spreadWind}
\end{figure}

\begin{figure}[H]
	\begin{center}
		\includegraphics[width=0.49\textwidth]{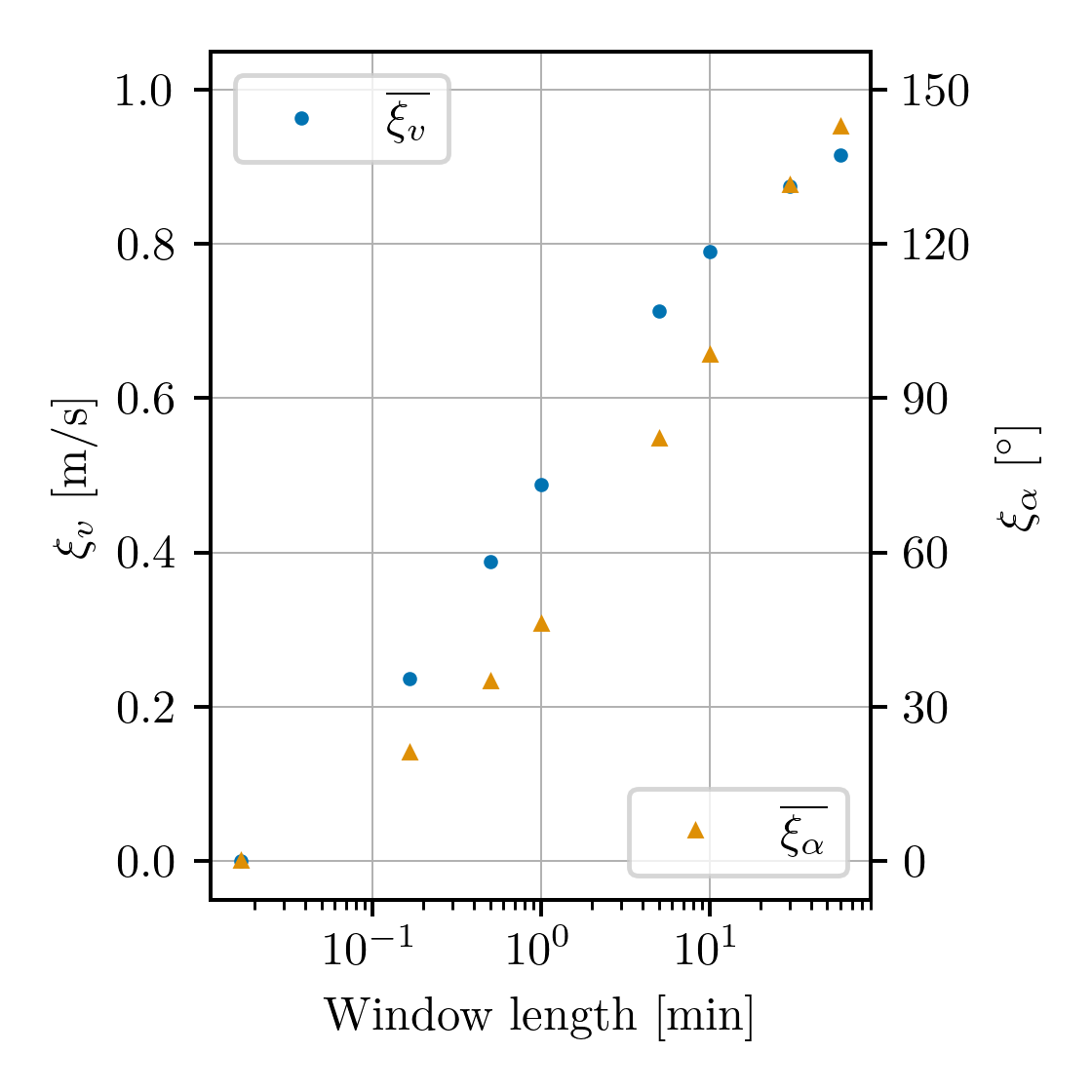}			
		\includegraphics[width=0.49\textwidth]{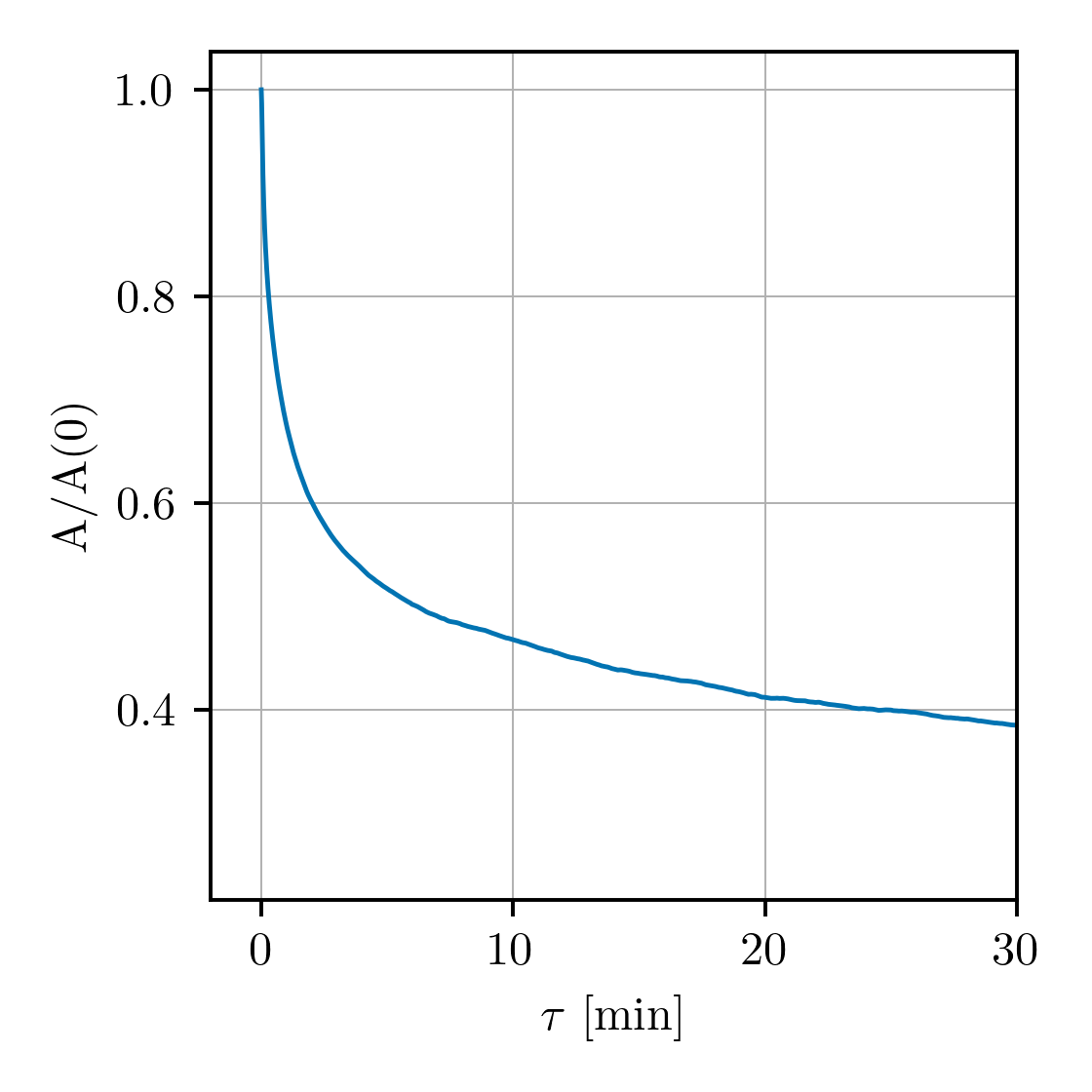}	
	\end{center}
	\caption{Left: Dependence of the mean variability and variability of the variability in wind speed and direction on the averaging window length. Right: Autocorrelation of \SI{1}{\second} wind data.}
	\label{fig:analysisWindow_spreadWindMean}
\end{figure}

We mentioned in Section~\ref{sec:wind-data-n-realistic-conditions} that the length of the averaging window affects the wind variability inside the window. Let us take a look at some numbers. Figure~\ref{fig:analysisWindow_spreadWind} shows the distribution of variabilities in wind speed (left) and direction (right) for all the observed data for different lengths of the window. To compute DTR accurately, the average speed and direction should be representative of that window, so ideally, a low variability within the window is needed, with low $\xi$ and narrow $\xi$ distributions. This is true for very short windows, for example the \SI{10}{\second} window. For both parameters, the distribution widens with the longer window, with the mean $\overline{\xi}$ shifting towards a higher variability value, which is shown quantitatively in Figure~\ref{fig:analysisWindow_spreadWindMean}~(left). Judging by the decay of distributions, a \SI{5}{\minute} averaging window is already quite long in the context of the observed data. Additionally, we can look at the autocorrelation of wind data. To compute it, we use this form with the scalar product
\begin{equation}
	A(\tau) = \sum \mathbf{v}(t_i)\cdot \mathbf{v}(t_i+\tau).
	\label{eq:autocorrelation}
\end{equation}
Figure~\ref{fig:analysisWindow_spreadWindMean}~(right) shows the autocorrelation for \SI{1}{\second} data. We can see that the decay is smooth, with the slope already flattening considerably at \SI{5}{\minute}. Looking at the decay rate, the half-time is \SI{6.2}{\minute} (although the exponential fit fails to satisfactorily represent the data), supporting that a \SI{5}{\minute} window is long for the characteristic winds at the observed location.

\begin{figure}[H]
	\begin{center}
		\includegraphics[width=0.49\textwidth]{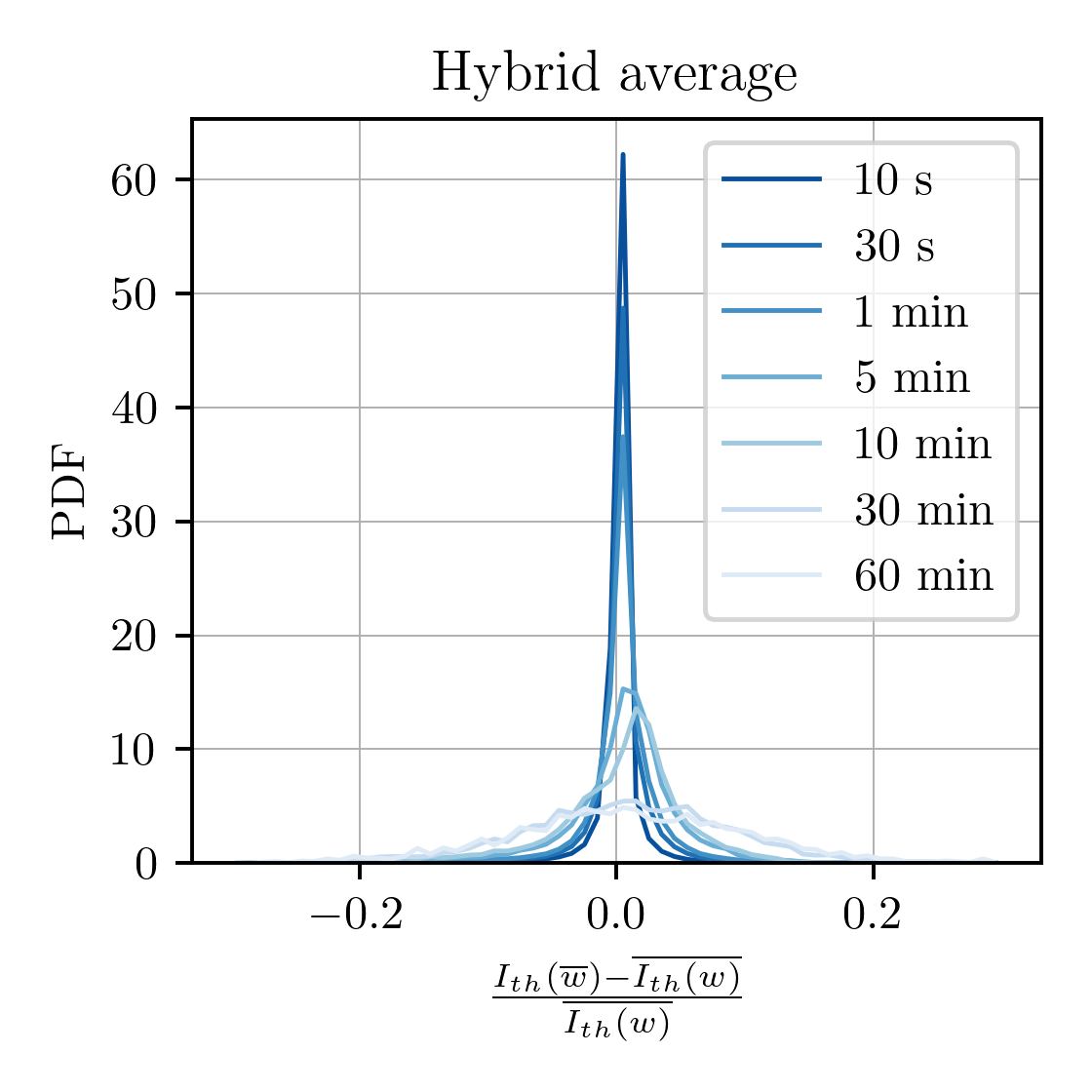}		
		\includegraphics[width=0.49\textwidth]{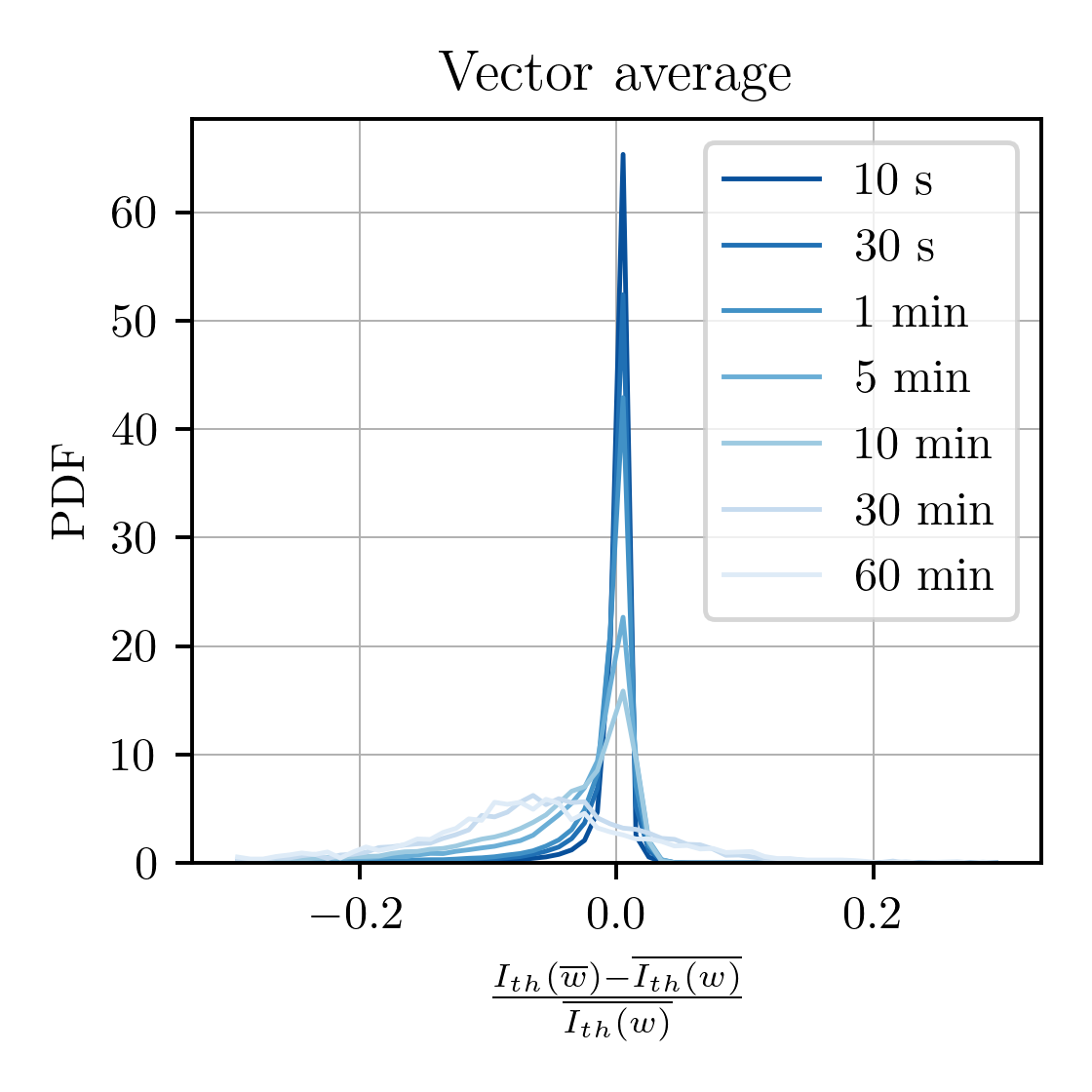}
	\end{center}
	\caption{Distribution of ampacity differences for different averaging window lengths for hybrid averaging (left) and vector averaging (right).}
	\label{fig:analysisWindow_errorHist}
\end{figure}

Next, let us take a look at the ampacity. Figure~\ref{fig:analysisWindow_errorHist} shows the relative ampacity differences distribution for different window lengths for both averaging methods. Again, the distributions of ampacity differences get wider with the increasing window length. Additionally, for vector averaging, the distributions noticeably shift to the left with the longer window. Thus the averaged-data ampacity underestimates the \Iavg, which is expected, as we saw that a longer window causes larger variability in wind direction $\xi_{\alpha}$, which lowers the vector-averaged wind speed (amplitude). The \SI{1}{\minute} window with standard deviation $\sigma_{hybrid}=0.031$ and $\sigma_{vector}=0.036$ is a considerable improvement over the \SI{5}{\minute} window ($\sigma_{hybrid}=0.046$, $\sigma_{vector}=0.058$) while choosing a longer \SI{10}{\minute} window ($\sigma_{hybrid}=0.052$, $\sigma_{vector}=0.064$) does not have a huge effect. This is especially true for the hybrid averaging, where the distribution is very similar to the one for the \SI{5}{\minute} window for the observed data and the characteristic wind at the observed location. Lengthening the window to half an hour and more, on the other hand, significantly increases the differences.

One more thing to consider about different windows is that the wind measurements are used for the calculations of the ampacity over the whole span, which means the variability should be representative of that span. At different window lengths, this would not necessarily be the case, so the span might have to be divided into several sub-spans. Additional data with multiple wind measurement sites would be needed to determine where this transition happens, so this is an opportunity for future research.

\section*{Acknowledgements}
Authors acknowledge the financial support from the Slovenian Research And Innovation Agency (ARIS) research core funding No.\ P2-0095, and the HEDGE-IoT project, which has received funding from the European Union's Horizon Europe research and innovation programme under grant agreement No. 101136216. We would also like to express our gratitude to the Slovenian TSO, ELES, Ltd., for providing the data required for this study.

\bibliographystyle{ieeetr}
\bibliography{refs}

\begin{thebibliography}{10}

\bibitem{bichler2022electricity}
M.~Bichler, H.~U. Buhl, J.~Kn{\"o}rr, F.~Maldonado, P.~Schott, S.~Waldherr, and
  M.~Weibelzahl, ``Electricity markets in a time of change: a call to arms for
  business research,'' {\em Schmalenbach Journal of Business Research},
  vol.~74, no.~1, pp.~77--102, 2022.

\bibitem{cimini2013temperature}
C.~A. Cimini~Jr and B.~Q.~A. Fonseca, ``Temperature profile of progressive
  damaged overhead electrical conductors,'' {\em International Journal of
  Electrical Power \& Energy Systems}, vol.~49, pp.~280--286, 2013.

\bibitem{a_douglass_review_2019}
D.~A.~Douglass, I.~Grant, J.~A. Jardini, R.~Kluge, P.~Traynor, C.~Davis,
  J.~Gentle, H.-M. Nguyen, W.~Chisholm, C.~Xu, T.~Goodwin, H.~Chen,
  S.~Nuthalapati, and N.~Hurst, ``A {Review} of {Dynamic} {Thermal} {Line}
  {Rating} {Methods} {With} {Forecasting},'' {\em IEEE Transactions on Power
  Delivery}, vol.~34, pp.~2100--2109, Dec. 2019.

\bibitem{maksic_cooling_2019}
M.~Maksić, V.~Djurica, A.~Souvent, J.~Slak, M.~Depolli, and G.~Kosec,
  ``Cooling of overhead power lines due to the natural convection,'' {\em
  International Journal of Electrical Power \& Energy Systems}, vol.~113,
  pp.~333--343, 2019.
\newblock Publisher: Elsevier.

\bibitem{cigre}
``Guide for thermal rating calculations of overhead lines,'' guide, CIGRE,
  2014.

\bibitem{ieee}
``Ieee standard for calculating the current-temperature relationship of bare
  overhead conductors, ieee std 738™-2023,'' standard, IEEE, 2023.

\bibitem{irena_dynamic_2020}
``Dynamic line rating - innovation landscape brief,'' 2020.

\bibitem{matus2012identification}
M.~Matus, D.~S{\'a}ez, M.~Favley, C.~Suazo-Mart{\'\i}nez, J.~Moya,
  G.~Jim{\'e}nez-Est{\'e}vez, R.~Palma-Behnke, G.~Olgu{\'\i}n, and P.~Jorquera,
  ``Identification of critical spans for monitoring systems in dynamic thermal
  rating,'' {\em IEEE Transactions on Power Delivery}, vol.~27, no.~2,
  pp.~1002--1009, 2012.

\bibitem{karimi_2018_dynamic}
S.~Karimi, P.~Musilek, and A.~M. Knight, ``Dynamic thermal rating of
  transmission lines: A review,'' {\em Renewable and Sustainable Energy
  Reviews}, vol.~91, pp.~600--612, 2018.

\bibitem{poli_possible_2019}
D.~Poli, P.~Pelacchi, G.~Lutzemberger, T.~{Baffa Scirocco}, F.~Bassi, and
  G.~Bruno, ``The possible impact of weather uncertainty on the dynamic thermal
  rating of transmission power lines: A monte carlo error-based approach,''
  {\em Electric Power Systems Research}, vol.~170, pp.~338--347, 2019.

\bibitem{rashkovska_uncertyinty_2022}
A.~Rashkovska, M.~Jančič, M.~Depolli, J.~Kosmač, and G.~Kosec, ``Uncertainty
  assessment of dynamic thermal line rating for operational use at transmission
  system operators,'' {\em IEEE Transactions on Power Systems}, vol.~37, no.~6,
  pp.~4642--4650, 2022.

\bibitem{chen_secure_2024}
Z.~Chen, B.~Zhang, C.~Du, P.~Li, and W.~Meng, ``Secure probabilistic interval
  prediction of dynamic thermal rating against weather imbalance constraints,''
  {\em IEEE Transactions on Industrial Informatics}, vol.~20, no.~10,
  pp.~12375--12384, 2024.

\bibitem{pytlak_modelling_2011}
P.~Pytlak, P.~Musilek, E.~Lozowski, and J.~Toth, ``Modelling precipitation
  cooling of overhead conductors,'' {\em Electric Power Systems Research},
  vol.~81, no.~12, pp.~2147--2154, 2011.

\bibitem{kosec_dynamic_2017}
G.~Kosec, M.~Maksić, and V.~Djurica, ``Dynamic thermal rating of power lines
  -- model and measurements in rainy conditions,'' {\em International Journal
  of Electrical Power \& Energy Systems}, vol.~91, pp.~222--229, 2017.

\bibitem{hosek_effect_2011}
J.~Hosek, P.~Musilek, E.~Lozowski, and P.~Pytlak, ``Effect of time resolution
  of meteorological inputs on dynamic thermal rating calculations,'' {\em IET
  Generation, Transmission \& Distribution}, vol.~5, no.~9, p.~941, 2011.

\bibitem{howington_dynamic_1987}
B.~S. Howington and G.~J. Ramon, ``Dynamic thermal line rating summary and
  status of the state-of-the-art technology,'' {\em IEEE Transactions on Power
  Delivery}, vol.~2, no.~3, pp.~851--858, 1987.

\bibitem{suomi_wind_2018}
I.~Suomi and T.~Vihma, ``Wind gust measurement techniques—from traditional
  anemometry to new possibilities,'' {\em Sensors}, vol.~18, no.~4, 2018.

\bibitem{rafei_analysis_2023}
M.~El~Rafei, S.~Sherwood, J.~Evans, and A.~Dowdy, ``Analysis and
  characterisation of extreme wind gust hazards in new south wales,
  australia,'' {\em Natural Hazards}, vol.~117, pp.~1--21, 03 2023.

\bibitem{raichle2009wind}
B.~W. Raichle and W.~R. Carson, ``Wind resource assessment of the southern
  appalachian ridges in the southeastern united states,'' {\em Renewable and
  Sustainable Energy Reviews}, vol.~13, no.~5, pp.~1104--1110, 2009.

\bibitem{rott_robust_2018}
A.~Rott, B.~Doekemeijer, J.~K. Seifert, J.-W. van Wingerden, and M.~K\"uhn,
  ``Robust active wake control in consideration of wind direction variability
  and uncertainty,'' {\em Wind Energy Science}, vol.~3, no.~2, pp.~869--882,
  2018.

\bibitem{mahrt_surface_2011}
L.~Mahrt, ``Surface wind direction variability,'' {\em Journal of Applied
  Meteorology and Climatology}, vol.~50, no.~1, pp.~144 -- 152, 2011.

\bibitem{shu_analysis_2024}
Z.~Shu, P.~Chan, and X.~He, ``Analysis of horizontal wind direction variability
  considering different influencing factors,'' {\em Journal of Wind Engineering
  and Industrial Aerodynamics}, vol.~252, p.~105819, 2024.

\bibitem{houle_near_2023}
J.~Houle and F.~van Breugel, ``Near-surface wind variability over
  spatiotemporal scales relevant to plume tracking insects,'' {\em Physics of
  Fluids}, vol.~35, p.~055145, 05 2023.

\bibitem{meteo}
``Guide to instruments and methods of observation, volume i - measurement of
  meteorological variables,'' guide, World Meteorological Organization, 2023.

\bibitem{gilhousen_field_1987}
D.~B. Gilhousen, ``A field evaluation of ndbc moored buoy winds,'' {\em Journal
  of Atmospheric and Oceanic Technology}, vol.~4, no.~1, pp.~94 -- 104, 1987.

\bibitem{thomas_buoy_2011}
B.~R. Thomas and V.~R. Swail, ``Buoy wind inhomogeneities related to averaging
  method and anemometer type: application to long time series,'' {\em
  International Journal of Climatology}, vol.~31, no.~7, pp.~1040--1055.

\bibitem{ward_time-averaged_2023}
K.~R. Ward, O.~Bamisile, C.~J. Ejiyi, and I.~Staffell, ``Time-averaged wind
  power data hides variability critical to renewables integration,'' {\em
  Energy Strategy Reviews}, vol.~50, p.~101235, Nov. 2023.

\bibitem{gonzalez-cagigal_influence_2022}
M.~González-Cagigal, J.~Rosendo-Macías, A.~Bachiller-Soler, and
  D.~Señas-Sanvicente, ``Influence of the wind variability on the calculation
  of dynamic line rating,'' {\em Electric Power Systems Research}, vol.~211,
  p.~108234, Oct. 2022.

\bibitem{tuller-brett_characteristics_1984}
S.~E. Tuller and A.~C. Brett, ``The characteristics of wind velocity that favor
  the fitting of a weibull distribution in wind speed analysis,'' {\em Journal
  of Applied Meteorology and Climatology}, vol.~23, pp.~124--134, 1984.

\bibitem{copernicus_how_nodate}
``How to average winds? {\textbar} {Copernicus} {Marine} {Help} {Center}.''

\bibitem{lee2018assessing}
J.~C. Lee, M.~J. Fields, and J.~K. Lundquist, ``Assessing variability of wind
  speed: comparison and validation of 27 methodologies,'' {\em Wind Energy
  Science}, vol.~3, no.~2, pp.~845--868, 2018.

\bibitem{arenasLopez2020stochastic}
J.~P. Arenas-López and M.~Badaoui, ``Stochastic modelling of wind speeds based
  on turbulence intensity,'' {\em Renewable Energy}, vol.~155, pp.~10--22,
  2020.

\bibitem{morgan_rating_1967}
V.~Morgan, ``Rating of bare overhead conductors for continuous currents,'' {\em
  Proceedings of the Institution of Electrical Engineers}, vol.~114,
  pp.~1473--1482, 1967.

\bibitem{morgan_heat_1973}
V.~Morgan, ``The heat transfer from bare stranded conductors by natural and
  forced convection in air,'' {\em International Journal of Heat and Mass
  Transfer}, vol.~16, no.~11, pp.~2023--2034, 1973.

\bibitem{waghorne_current_1951}
J.~H. Waghorne and V.~E. Ogorodnikov, ``Current carrying capacity of acsr
  conductors,'' {\em Transactions of the American Institute of Electrical
  Engineers}, vol.~70, no.~2, pp.~1159--1162, 1951.

\bibitem{fand_continuous_1972}
R.~Fand and K.~Keswani, ``A continuous correlation equation for heat transfer
  from cylinders to air in crossflow for reynolds numbers from 10 to -2 to 2 x
  10 to 5,'' {\em International Journal of Heat and Mass Transfer}, vol.~15,
  no.~3, pp.~559--562, 1972.

\bibitem{morgan_thermal_1982}
V.~Morgan, ``The thermal rating of overhead-line conductors part i. the
  steady-state thermal model,'' {\em Electric Power Systems Research}, vol.~5,
  no.~2, pp.~119--139, 1982.

\end{thebibliography}

\end{document}